\documentclass{article}
\usepackage{graphicx} 
\usepackage[margin=1in]{geometry}
\usepackage[T1]{fontenc}
\usepackage[backend=bibtex, sorting=none, url=false, giveninits=true]{biblatex}
\usepackage{hyperref}
\usepackage{etoolbox}
\usepackage{stix}
\usepackage{tikz}
\usetikzlibrary{shapes.geometric}
\usepackage{amsmath}
\usepackage{amsfonts}
\usepackage{mathtools}
\usepackage{subfiles}
\usepackage[export]{adjustbox}
\usepackage{ragged2e}
\usepackage{bm} 
\usepackage[caption=false]{subfig}
\usepackage[normalem]{ulem}
\usepackage{graphicx}
\usepackage{wrapfig}
\usepackage{sidecap}
\usepackage{algorithm}
\usepackage{algpseudocode}
\usepackage{pdfpages}

\usepackage[noblocks]{authblk}

\newcommand{\dt}{{\Delta t}}

\newcommand{\s}{\sigma}
\renewcommand{\a}{\alpha}
\renewcommand{\b}{\beta}
\newcommand{\ddt}{\partial_t}

\usepackage{mathtools}

\title{A practical investigation on time integration in the quantized tensor train format}
\author[a]{Erika Ye}
\date{\today}

\affil[a]{\small{Applied Mathematics and Computational Research Division, Lawrence Berkeley National Laboratory, 1 Cyclotron Rd, Berkeley, CA 94720}}

\begin{document}
\maketitle

\begin{abstract}

    Quantized tensor trains (QTTs) are a multiscale computational framework that can potentially reduce the computational cost of solving partial differential equations and initial value problems by making low-rank approximations. However, its use is somewhat limited in practice, 
    in part due to the challenges that arise when making low-rank approximations of the quantized data. 
    For example, when performing long-time dynamical numerical simulations, it has been observed that the accumulation of numerical errors arising from both the discretization of the partial differential equation itself and the low-rank approximation can lead to increased rank and noise-dominated results.
    Focusing on a set of advection-dominated test problems relevant to electromagnetic plasmas and electromagnetic fields, this work investigates how the choice in time integrator, the addition of numerical dissipation, and the choice in problem representation can affect the efficiency and success of the QTT calculations. 
\end{abstract}

\maketitle

\section{Introduction}

Tensor trains are a low-rank representation that have demonstrated significant utility in a variety of numerical simulations, demonstrating significant reduction of computational cost for various problems in numerical simulation, including solving for the ground state or low-lying excited states in quantum chemistry applications \cite{White1993density, White2005density, Chan2002highly, chan2016matrix, dolgov_computation_2014}, solving linear problems \cite{Oseledets2012solution, Dolgov2014alternating}, and performing time integration to solve initial value problems in various scientific domains \cite{einkemmer_review_2024, einkemmer_low-rank_2018, christlieb_sampling_2025, zheng_SLAR_2025}. The focus of this paper is on low-rank quantized tensor train (QTT) time integrators, which have been tested on electromagnetic simulations~\cite{manzini_tensor_2023, nguyen2026tensor}, the chemical master equation~\cite{kazeev_direct_2014}, Navier-Stokes calculations of wall-bounded flows~\cite{kiffner_tensor_2023, kornev_numerical_2023, peddinti_complete_2023} and in turbulent regimes ~\cite{Gourianov2021, gourianov2022exploiting, hulst_quantum-inspired_2026}, kinetic plasmas~\cite{ye_quantum-inspired_2022, ye_quantized_2024}, and more.

While all low-rank solvers face the uncertainty of whether the representation of choice can accurately represent the final solution, low-rank time integrators face additional challenges. In particular, even if the solution at the final time should be of modest rank, there is no guarantee that the solution at intermediate times is also low rank. Additionally, the accumulation of rank truncation errors and other numerical errors at each time step may corrupt the final result, yielding an unexpectedly high-rank result. This is especially concerning for advection-dominated systems in which there is little dissipation to help smooth out these errors. Understanding how to mitigate these issues is critical for practical usage of low-rank solvers for modeling dynamical systems. 

In contrast to non-quantized or functional tensor trains (TTs), which decompose high-dimensional data according to the dimensions of the problem, QTTs decompose data between the different length scales of a dyadic grid. As a result, it has been hypothesized that this representation may be beneficial for systems with some degree of scale separation, such that the behavior at fine scales is only weakly coupled to the behavior at coarse scales. One target application is to effectively capture Kolmogorov turbulence, in which energy at large scales is cascaded to finer scales
\cite{Gourianov2021, hulst_quantum-inspired_2026}.
%
However, before modeling turbulent dynamics, it is prudent to understand the limitations of QTT time integration for non-turbulent dynamics, especially to ensure that the simulation of systems that are expected to be low-rank remain low-rank over long simulation times. This would be also be important for efficiently capturing the dynamics before the onset of turbulence. 
In such scenarios, while the QTT implementation may not outperform functional TT methods or be worthwhile over standard numerical methods given the approximation error and computational overhead, its performance should still be reasonable. This paper is an exploration of the limitations of time integration in the QTT framework, to better understand when it performs poorly and how any problematic behavior can be overcome.

This paper is organized as follows. The first section provides a brief overview of QTTs and time integration algorithms. This is followed by three advection-dominated numerical examples requiring both small and large QTT ranks. In these examples, multiple approaches to maintain modest ranks throughout the time integration procedure are tested. Finally, the paper concludes with a discussion on the observations.

\section{Overview of quantized tensor trains}



Constructing a QTT involves first defining a quantization procedure to map a vector to a high-dimensional tensor space, and then using the tensor train decomposition to obtain a low-rank representation.  
This section provides an overview of the QTT ansatz for one-dimensional data in real space discretized on a uniform grid, largely following the discussions presented in Refs.~\cite{Khoromskij2011, Lindsey2023multiscale, Ripoll2021quantuminspired}. With this definition, the QTT is a multiscale ansatz that allows one to take advantage of any low-rank structure of the data between different length scales.

Let $x \in [0,1)$ denote the spatial variable. In the quantization procedure, $x$ is mapped to an infinite set of hyperindices $\{\vec{\sigma}\}$ = $\{ (\sigma_1, \sigma_2, \hdots) \}$. One straightforward choice is the binary decimal expansion of $x$:
\begin{align}
    x = \sum_{k=1}^\infty 2^{-k} \sigma_k = 0.\sigma_1 \sigma_2 \sigma_3 \hdots
\end{align}
When the decimal expansion is truncated to depth $L$, the hyperindices map to a uniformly discretized grid $\mathcal{D}_L$ with grid spacing $\Delta x=1/2^L$ and grid points $x_n = n/2^L$, where $n \in \{0, 1, ..., N-1\}$.
The $k^\text{th}$ element in the hyperindex $\vec{\sigma}$ corresponds to the $k^\text{th}$ grid scale, and the grids at different scales are separable. For example, partitioning the grid at depth $m$ yields
\begin{align}
    x_{\leq m} := \sum_{k=1}^m 2^{-k} \sigma_k, 
    \qquad
    x_{>m} := \sum_{k=m+1}^L 2^{-k} \sigma_k = 2^{-m} \sum_{k=1}^{L-m} 2^{-k} \sigma_k.
\end{align}
where the grid points $x_{\leq m} \in \mathcal{D}_m$ form a coarser grid, while $x_{>m} \in 2^{-m} \mathcal{D}_{L-m}$ is the fine scale grid as it has been scaled by $2^{-m}$.

On the partitioned grids, a function $f(x)$ can be decomposed as
\begin{align}
   f(x) = f(x_{\leq m} + x_{>m}) =  
   \sum_\alpha c_\alpha \psi_\alpha(x_{\leq m}) \,  \varphi_\alpha(x_{>m}),
\end{align}
similar to a modal discontinuous Galerkin representation where $\varphi_\alpha(x_{>m})$ is the basis function within each cell of the coarse grid $\mathcal{D}_m$. However, note that in this expression, the basis functions are discretized.


In the QTT representation of $f(x)$, the above decomposition is performed at every grid scale in an iterative fashion. Starting from the quantized representation of $f(x)$, which is the $L$-dimensional tensor $f(\s_1, \s_2, \hdots,\s_L) $, performing the standard tensor train decomposition based on the iterative application of the singular value decomposition (SVD)~\cite{Oseledets2011} yields the quantized tensor train
\begin{align}
    f(x_n) = f(\s_1, \s_2, ..., \s_L) & = 
    \sum_{\alpha_1=1}^{r_1} ... \sum_{\alpha_{L-1}=1}^{r_{L-1}} 
    M^{(1)}_{\alpha_1} (\s_1) M^{(2)}_{\alpha_1, \alpha_2}(\s_2) \hdots M^{(L)}_{\alpha_{L-1}} (\s_L) 
    \label{eq:qtt_ansatz}
\end{align}
where $M^{(k)} \in \mathbb{C}^{2\times r_{k-1} \times r_k} $ are 3-dimensional tensor cores with one index ($\s_k$) corresponding to a physical dimension and two indices ($\a_{k-1},\a_k$) corresponding to virtual dimensions (though $r_0=r_K=1$ so the indices $\alpha_0$ and $\alpha_K$ are dropped). 
The rank of the tensor train is defined as $\text{max}(r_1, \hdots, r_{L-1})$, which can have a maximum value of $2^{\lfloor L/2 \rfloor}$.

Linear operators $A$ acting on the vector data can also be written in QTT form,
\begin{align}
    A(x_n', x_n) & = T((o_1,\hdots,o_L), (i_1,\hdots, i_L))
    = \sum_{\beta_1 = 1}^{r_1} \hdots \sum_{\beta_{L-1}}^{r_{L-1}} M_{\b_1}^{(1)}(o_1,i_1) M_{\b_1,\b_2}^{(2)} (o_2, i_2) \hdots M^{(L)}_{\b_{L-1}} (o_L, i_L)
    \label{eq:qtt_matrix}
\end{align}
where now $M^{(k)} \in \mathbb{C}^{2\times 2 \times r_{k-1} \times r_k} $ are 4-dimensional tensor cores, with $i_k$ and $o_k$ indicating the physical input and output dimensions of the operator, respectively.
It has been shown that sparse and structured operators, such as Toeplitz matrices and circulant matrices, have low-rank representations \cite{Khoromskij2011, kazeev_multilevel_2013}. 

Basic linear operations such as vector-vector addition and matrix-vector multiplication can be performed efficiently while remaining in the QTT format using standard TT algorithms~\cite{schollwock_density-matrix_2011}. However, these operations increase the rank of the TT being operated on, and a rank truncation procedure is needed to return the rank to a manageable size. Standard rank truncation schemes are based on the SVD, in which one can choose to retain the $r_{max}$ largest singular values, or retain only the singular values larger than some threshold value $\varepsilon$ after normalization with respect to the $L_2$ norm~\cite{Oseledets2011}. In the naive implementation, the computational cost scales like $\mathcal{O}(r_A^3r^3)$, where $r_A$ is the rank of the time evolution operator and $r$ is the rank of the original TT. Fortunately, approximate methods, like the zip-up algorithm \cite{stoudenmire_minimally_2010} (used in this work) and randomized methods \cite{camano_successive_2025}, can reduce the cost to $\mathcal{O}(r_A r\, \tilde{r}^2)$, where $\tilde{r}$ is the TT rank of the true solution, and $\tilde{r} \leq r_A r$. 
In the worst case, these methods would offer no advantage over a naive implementation. Fortunately, in practice, moderate speed-up is often observed, especially if larger errors can be tolerated.

\subsection{Mappings for multi-dimensional problems}
\label{sec:mapping}
Extending the above one-dimensional quantization to $K$-dimensional multivariate data is straightforward. The variable corresponding to the $i^\text{th}$ dimension, $x_i$, is mapped to its corresponding hyperindex $(\s_{i1}, \s_{i2}, \hdots \s_{iL_i})$, where $L_i$ is the depth or resolution of the grid along the corresponding dimension. Following the procedure above, the result is a tensor train with $\sum_i L_i$ tensor cores. However, there are multiple options in ordering the hyperindices. 
One could consider an interleaved ordering in which the hyperindices are grouped by grid scale:
\begin{align}
    f(x_1, x_2, \hdots, x_K) = T( 
    \underbrace{\s_{11}, \s_{21}, \hdots, \s_{L1}}_{\text{depth 1}},
    \underbrace{\s_{12}, \s_{22}, \hdots, \s_{L2}}_{\text{depth 2}}, \hdots, 
    \underbrace{\s_{1K}, \s_{2K}, \hdots, \s_{LK}}_{\text{depth } L}, )
    \label{eq:int_qtt}
\end{align}
where $L_i = L$ for all dimensions $i$ for simplicity. The indices at each grid resolution are often grouped together, $\tau_i = (\s_{1i}, \s_{2i}, \hdots, \s_{Li})$, yielding a tensor train with $L$ tensor cores.
Alternatively, the hyperindices for each dimension can be appended sequentially,
\begin{align}
    f(x_1, x_2, \hdots, x_K) = T( 
    \underbrace{\s_{11}, \s_{12}, \hdots, \s_{1L_1}}_{x_1},
    \underbrace{\s_{21}, \s_{22}, \hdots, \s_{2L_2}}_{x_2}, \hdots, 
    \underbrace{\s_{K1}, \s_{K2}, \hdots, \s_{KL_K}}_{x_K}, ).
    \label{eq:seq_qtt}
\end{align}
In addition to the ordering of the dimensions, there is also freedom in the directionality of the indices as they can be arranged in a forwards orientation $(\sigma_1, \hdots, \sigma_L)$ or a backwards orientation $(\sigma_L, \hdots, \sigma_1)$.

In the rest of the paper, ``int'' will be used to denote the interleaved mapping with grouping (Eq.~\eqref{eq:int_qtt}), while, ``seq'' will be used to denote the sequential ordering (Eq.~\eqref{eq:seq_qtt}). For sequentially ordered QTTs, ``F'' and ``B'' will be used to further specify if the forward or backwards tensor train orientation for each of the dimensions. 
For example, for $K=2$, the mappings used in this work include
\begin{itemize}
    \item ``seq(FB)'', with indices ordered like $(\s_{11}, \s_{12}, \hdots, \s_{1L_1}, \s_{2L_2}, \hdots, \s_{22}, \s_{21} )$
    \item ``seq(FF)'', with indices ordered like $(\s_{11}, \s_{12}, \hdots, \s_{1L_1}, \s_{21}, \s_{22}, \hdots, \s_{2L_{2}})$
    \item ``seq(BF)'', with indices ordered like $(\s_{1L_1}, \hdots  \s_{12}, \s_{11}, \s_{21}, \s_{22}, \hdots, \s_{2L_2} )$
\end{itemize}


Aside from the tensor train, one could consider other tensor network ansatz\"{e}, such as a tree tensor network ~\cite{ye_quantized_2024, dolgov_two-level_2013} or even a 2-D tensor network (PEPS). However, this is beyond the scope of this paper.

\subsection{Time integration}
In this paper, a variety of time integrators are used. They largely fall under three categories: step-and-truncate (SAT) methods which best suited for explicit integrators, alternating optimization methods which are better suited for implicit integrators, and quantized dynamical low-rank (qDLR)-type methods which can be used for both. This section will offer a brief summary of each. Further details can be found in the referenced papers. 

The idea behind step-and-truncate (SAT) methods is to first compute the next time step exactly in the TT format before performing a rank truncation procedure. This is possible because many classical time integrators, including explicit integrators and implicit integrators that can be solved using gradient descent~\cite{fraschini_symplectic_2024}, only require matrix-vector multiplication and vector addition operations. As discussed in the previous section, these operations can be performed efficiently in the TT format, though a rank truncation procedure must follow because the operations yield a QTT of increased rank. Thus, SAT allows one to obtain a near-optimal low-rank representation of the next time step. However, SAT can quickly become very expensive, especially if a single time step requires performing multiple operations.
%
%
In this paper, the explicit time integrators implemented with SAT include the finite difference time domain (FDTD) method and the first-order finite volume (FV) method for solving Maxwell's equation, fourth-order Runge-Kutta (RK4), and Lax-Wendroff with the Maccormack method (LW). For RK4, SVD-based rank truncation is performed at each intermediate stage; otherwise the computation becomes very slow. In contrast, for the Maccormack method, rank truncation is performed only after the second and final stage (details are in the Supplementary Information (SI) Section 1).

Implicit time integration can also be performed using alternating tensor train optimization algorithms based on alternating least squares (ALS) \cite{Oseledets2012solution} or inexact gradient-based minimization of the residual~\cite{Dolgov2014alternating}. Similar algorithms based on cross interpolation \cite{Oseledets2010cross} can be used to efficiently perform nonlinear operations as well \cite{gharemani_cross_2024}. In this paper, the ALS tensor train solver is used when performing Crank-Nicolson (CN) time integration. 

The remaining time integration schemes considered in this work follow the dynamical low rank (DLR) procedure with projector splitting
~\cite{koch_dynamical_2007, lubich_projector-splitting_2014, hochbruck_rank-adaptive_2023, einkemmer_low-rank_2018, ceruti_robust_2024, ceruti_galerkin_2025}. In DLR, each tensor core in the TT is evolved in time one by one, starting from one end of the TT and progressing sequentially to the other. The tensor core is evolved according to the original equation of motion after being projected onto the TT manifold, the reduced space defined by the tensor cores that are not being updated. These reduced equations of motion for each tensor core can be solved using any suitable time integrator. After the time evolution of the tensor core, the updated core is used to define the TT manifold for the time evolution of the next tensor core. However, once the manifold is updated and before the next tensor core is evolved forward in time, the TT itself must be evolved backwards in time to the original time step. How this is performed leads to two DLR variants. First is the original formulation, in which the backwards time evolution is computed directly (DLR-PS). This can be unstable for dissipative systems unless an implicit integrator is used \cite{lubich_projector-splitting_2014}. Second is the alternating basis-update projection variant in which the original state is projected onto the updated manifold (DLR-AP)~\cite{ceruti_galerkin_2025}. 
While often discussed in the matrix setting (i.e., a tensor train of length two) for non-quantized representations, DLR can readily be extended to the QTT setting in a straightforward manner, resulting in quantized DLR (qDLR)~\cite{ye_quantized_2024, ye_time_2025}. The same methods are also used in the quantum chemistry and condensed matter communities, with the two variants referred to as the time-dependent variational principle (TDVP) and time-dependent density matrix renormalization group (TD-DMRG), respectively \cite{feiguin2005time, Haegeman_TDVP, Paeckal_2019_time}. 

One caveat to DLR methods is that unlike SAT methods, the accuracy of the time evolution heavily depends on the ranks of the TT manifold and its ability to capture the full dynamics. As a result, several DLR variants that allow one to increase the TT rank before or during the DLR procedure have been developed. For DLR-PS, a two-site algorithm~\cite{Haegeman_TDVP} allows the TT rank to increase during the DLR procedure, with the rank increase determined by a maximum rank or a rank truncation threshold value. When the two-site algorithm also fails, one can augment the TT manifold with the time derivative before the DLR sweep~\cite{yang2020time}. This ensures that the dynamics of the initial state can be captured during the DLR procedure. For DLR-AP, rank expansion is performed during the basis update and projector procedure. Basis expansion is actually critical for improved accuracy for DLR-AP. If performed correctly, one can even achieve high-order accuracy with respect to time step size~\cite{feiguin2005time, nobile_robust_2025, ye_time_2025}. This is in contrast to qDLR-PS which in practice is only first or second-order accurate with respect to time step size depending on the projector splitting procedure.


%
The cost of these algorithms are dominated by the SVD, scaling like $\mathcal{O}(d\bar{r}^3)$, the tensor contraction, scaling like $\mathcal{O}(dr_A \bar{r}^3)$, and solving the reduced dynamical equation, with the vector size of $d\bar{r}^2$. Here, $d$ is the size of the physical dimension of the tensor core, and $\bar{r}$ is the the rank of the tensor train manifold on which the dynamics is evolved, which may be larger than the initial TT if an adaptive scheme is used. If this rank increase is allowed, then rank truncation is performed after DLR to reduce the TT rank again.
In this paper, qDLR using RK4 and CN as core-level solvers are considered. (The MacCormack time integrator is not well-suited for these methods, as qDLR provides little computational advantage compared to SAT methods and is more difficult to implement correctly. In the author's opinion, standard DLR is most beneficial when one is interested in using more costly time integrators.)

\section{Numerical examples}


\subsection {Whistler wave in plasmas}
\label{sec:whistler}

Consider the Whistler wave, an electron wave that occurs in plasmas in the presence of an external magnetic field. When the wave amplitude is small, the dynamics is dominated by by wave propagation, so the state at some future time should only differ from the initial state in phase and possibly amplitude. therefore, the initial state and final state should be of the same rank. The goal of this section is to verify this is satisfied even after long simulation times.

The dynamics of a plasma in the presence of an electromagnetic field is described by the Vlasov-Maxwell equations, with coordinates in position space $\textbf{x}$ and velocity space $\textbf{v}$. Because the low-rank approximation may result in loss of positivity and cause numerical issues, we solve a set of modified equations,
\begin{align}
    & \ddt g_s + \textbf{v} \cdot \nabla_\textbf{x} g_s + \frac{q_s}{m_s} \left(\textbf{E} + \textbf{v} \times \textbf{B} \right) \nabla_\textbf{v} g_s = 0
    \label{eq:boltzmann}
    \\
    & \ddt \textbf{E} = c^2 \nabla \times \textbf{B} - \frac{1}{\varepsilon_0} \sum_s q_s \int_V \textbf{v}  \, g_s^* g_s \, d\textbf{v}
    \label{eq:E}
    \\
    & \ddt \textbf{B} = -\nabla \times \textbf{E}
    \label{eq:B} 
\end{align}
where $\textbf{E}$ is the electric field, $\textbf{B}$ is the magnetic field, and $g_s$ is the probability amplitude of particle species $s$, such that $g_s^* g_s=f_s$ is the probability distribution function used in the standard Vlasov equations. Additionally, $q_s$ and $m_s$ are the charge and mass of the particle species, $\varepsilon_0$ is the vacuum permittivity, and $c$ is the speed of light. Here, the only particle species considered are electrons and protons.

\begin{figure}
    \includegraphics[width=\linewidth, trim={0 1cm 0 0}]{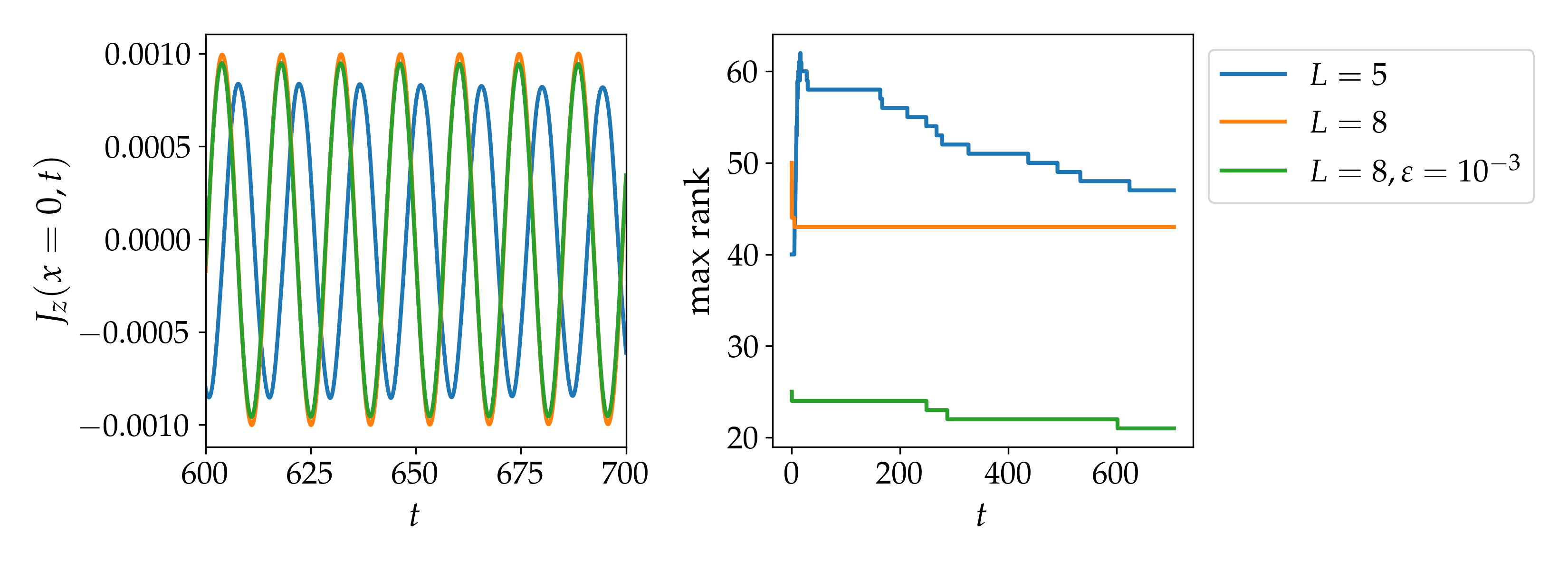}
    \caption{Ranks of $g_e$ for select Whistler wave simulations using MacCormack time integration with rank truncation threshold $\varepsilon = 2.47 \times 10^{-5}$. (left) The current density $J_z$ measured at $x=0$ over time. (right) The maximum ranks of the electron probability amplitude $g_e$ over time. The $L=8$ calculation uses a time step of $\Delta t=0.027$, which is larger than the CFL constraint. The $L=5$ calculation shows an error in phase and amplitude, resulting from the low resolution and the dissipative nature of the Lax-Wendroff scheme. Interestingly, increasing resolution requires a smaller rank. Utilizing a larger $\epsilon$ appears to increase error in the amplitude of the wave but not the frequency. The legend corresponds to runs labeled (2.1), (2.3), and (2.5) in Table~\ref{tab:whistler}. 
    }
    \label{fig:whistler_lw}
\end{figure}

\begin{figure}
    \includegraphics[width=\linewidth, trim={0 1cm 0 0}]{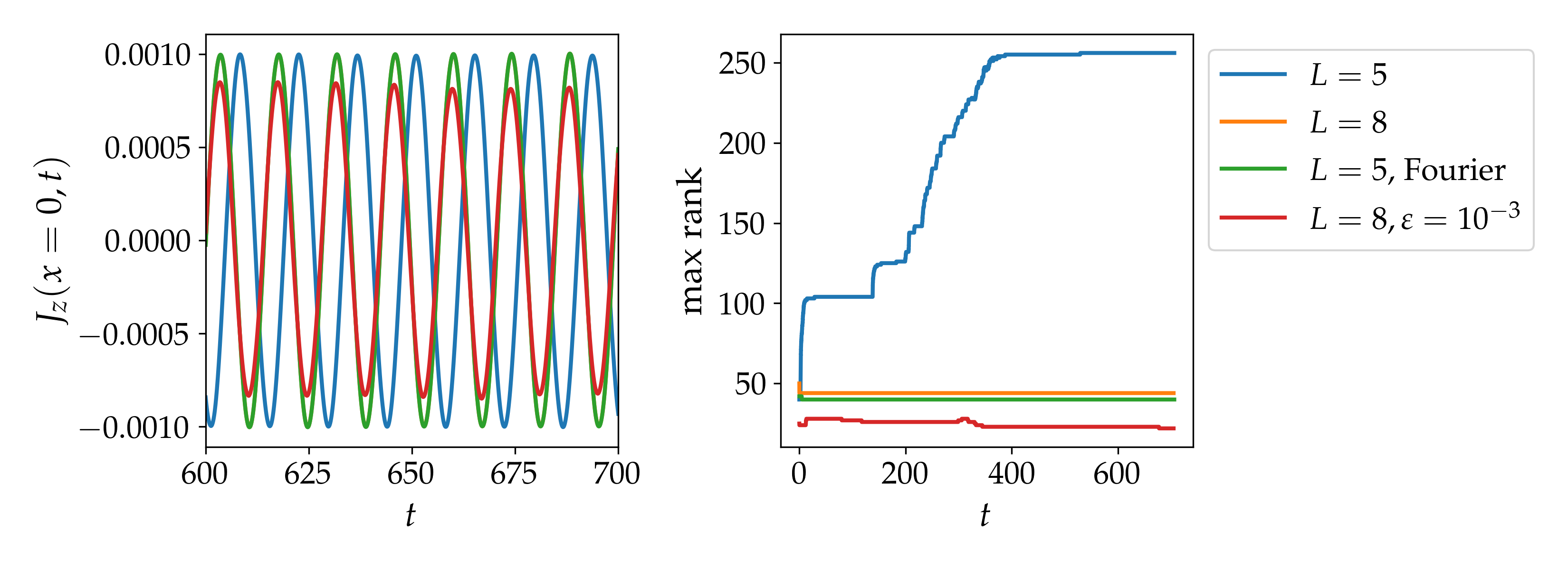}
    \caption{Same as Fig.~\ref{fig:whistler_lw} but for qDLR-PS + RK4 calculations. The $L=8$ calculations also use $\Delta t=0.27$. The finite difference $L=5$ calculation requires surprisingly large rank and exhibits a phase error due to the limited resolution. However, the $L=8$ calculation and the $L=5$ calculation in the Fourier basis yield similar dynamics and do not exhibit the unexpected growth in rank. Again, utilizing a larger $\epsilon$ appears to increase error in the amplitude of the wave but not the frequency. The legend corresponds to runs (3.1), (3.5), (3.8), and (3.7) in Table~\ref{tab:whistler}.
    }
    \label{fig:whistler_tdvp}
\end{figure}

The Whistler wave can be captured in a 1D3V (1 dimension in position space, 3 dimensions in velocity space) simulation. The initialization is not too important for the discussion and is described in Appendix~\ref{app:whistler}.  
The simulation is performed using $2^L$ grid points along each dimension. Unless otherwise stated, and a rank truncation threshold of about $\varepsilon = 2.47 \times 10^{-5}$ is used. (This value was determined by estimating the magnitude of the perturbation of the $E_z$ field and setting the error threshold to be less than that. 
However, it is observed that a larger threshold of $10^{-4}$ can also be tolerated.)
The QTT for the electron probability amplitude is constructed using a sequential mapping with a forward orientation for the spatial dimension and a backward orientation for the velocity dimensions, with the dimensions ordered like $(x,v_x,v_y,v_z)$. Maxwell's equations are solved using Crank-Nicolson, while the Boltzmann's equations for the electrons and are solved using the specified time integration schemes. The fixed-ion approximation is used, so the proton dynamics are ignored.
Unless otherwise specified a time step size based on the CFL condition is used.
For $L=5$, this corresponds to a time step size $\Delta t \approx 0.027 \, |\Omega^{-1}_{c,e}|$, where $\Omega_{c,e}$ is the electron cyclotron frequency. For $L=8$, this corresponds to $\Delta t \approx 0.00332|\Omega_{c,e}^{-1}|$. The units are chosen such that $\Omega_{c,e}=1$. 
The results are summarized in Table~\ref{tab:whistler}, and Figs.~\ref{fig:whistler_lw} and ~\ref{fig:whistler_tdvp}.  

First, fourth-order Runge-Kutta (RK4) was considered.
A fourth-order centered finite-difference stencil was used to approximate the spatial derivatives. With $L=5$, the rank of $g_e$ increased rapidly, reaching a value of about $r=400$ (run 1.1). The rank was even larger for a higher resolution calculation with $L=8$ (run 1.2). In these calculations, rank truncation was performed at each intermediate stage in the RK4 procedure in order to make the calculation more tractable. However, the accumulation was still observed when rank truncation was not performed at the intermediate stages.

Next, Lax-Wendroff (LW) time integration, implemented via the MacCormack method, was considered.
In contrast to RK4, the calculation with the LW method yielded ranks that increase slightly at first but then started to decay over simulation time, staying around a value of $r=40$ (run 2.1). This behavior is likely due to its dissipative nature of the time integrator. The effect of this dissipation can be observed in the decay of the amplitude of the propagating wave.  
Interestingly, for a higher resolution calculation with $L=8$ (run 2.2), the rank of $g_e$ remained controlled, and required an even smaller rank of around $r=35$. 
Separately, due to the stability of the time-stepping scheme, it was observed that one could still use a time step size of $\dt=0.027 |\Omega^{-1}_{c,e}|$ despite the increase in resolution (run 2.3). The time step size does not affect the rank of $g_e$. Increasing the rank truncation threshold (runs 2.4 and 2.5) appeared to only affect the amplitude of the wave, and not its frequency. The maximum ranks are also lower throughout the simulation time.


The next set of tests utilized adaptive qDLR with backwards time propagation (qDLR-PS), and RK4 was used to solve the reduced equations of motion for each tensor core. For $L=5$, the rank again increased in an unphysical manner, but saturated at a value of 256 (run 3.1). 
It appears that the cause of this increase in rank of $g_e$ for the qDLR + RK4 calculation was in fact due to numerical errors arising from the numerical discretization. 
When increasing the grid resolution to $L=8$ along each dimension, 
the rank no longer increased at long simulation times (run 3.2). As with the Lax-Wendroff case, a larger time step size ($\Delta t = 0.027 |\Omega_{c,e}^{-1}|$) could also be used, with little impact on the rank of $g_e$. The same conclusion was obtained by solving the problem using a Fourier representation~\cite{ye_time_2025} (assuming periodic boundary conditions in all directions, including the velocity dimensions) with $2^5$ modes per dimension (run 3.8). 
For both the $L=5$ and $L=8$ calculations, if one increased the rank threshold $\varepsilon$ (runs 3.2, 3.3, 3.6, and 3.7), the rank of $g_e$ was reduced, and the error mostly appeared in the amplitude of the wave. 

The last set of tests considered qDLR with alternating projection and basis expansion (qDLR-AP), again using RK4 to solve the reduced equations of motion at each tensor core.
In contrast to the SAT implementation of RK4, it was observed that qDLR-AP with RK4 bypassed the drastic rank increase with simulation time. Unfortunately, the accuracy of the dynamics, particularly in the amplitude of the waves, was worse than in the Lax-Wendroff and qDLR-PS case.
Higher resolution calculations (run 4.2) resulted in larger rank calculations, and utilizing a spectral representation (run 4.3) did not significantly alter the rank. For the $L=8$ calculation, there was a large spike in rank at early times, giving rise to a large maximum rank listed in Table~\ref{tab:whistler}, but the rank quickly dropped afterwards.



\begin{table}[]
    \centering
    \small
    \begin{tabular}{c|c|c|c|c|c|c|c}
         Run & Integrator & Resolution & Other params & Osc. freq. & Amplitude std. dev. & Max. rank & Final rank  \\ \hline
         (1.1) & RK4 * & L=5 & -- & 0.44753 & $2.3882 \times 10^{-6}$ & 414 & -- \\
         (1.2) & RK4 * & L=8 & $\varepsilon=10^{-4} $ & -- & -- & 693 & -- \\
         (2.1) & LW & L=5 & -- & 0.43618 & $4.2461 \times 10^{-5}$ & 62 & 47 \\
         (2.2) & LW & L=8 & -- & 0.44916 & $6.9174 \times 10^{-7}$ & 50 & 43 \\
         (2.3) & LW & L=8 & $\dt=0.027$ & 0.44495 & $2.034 \times 10^{-6}$ & 50 & 43 \\
         (2.4) & LW & L=8 & $\dt=0.027$, $\varepsilon=10^{-4}$ & 0.44495 & $2.14 \times 10^{-6}$ & 39 & 35 \\
         (2.5) & LW & L=8 & $\dt=0.027$, $\varepsilon=10^{-3}$ & 0.44495 & $1.431 \times 10^{-5}$ & 25 & 21 \\
         (3.1) & qDLR-PS + RK4 & L=5 & -- & 0.44495 & $3.338 \times 10^{-6}$ & 256 & 256 \\
         (3.2) & qDLR-PS + RK4 & $L=5$ & $\varepsilon=10^{-4}$ & 0.44496  & $2.979 \times 10^{-6}$ & 88 & 88\\
         (3.3) & qDLR-PS + RK4 & $L=5$ & $\varepsilon=10^{-3}$ & 0.44496  & $1.502 \times 10^{-5}$ & 30 & 19  \\
         (3.4) & qDLR-PS + RK4 & $L=8$ & -- & 0.44498 & $6.3835 \times 10^{-7}$ & 50 & 47 \\
         (3.5) & qDLR-PS + RK4 & $L=8$ &  $\Delta t=0.027$ & 0.44495 & $1.916 \times 10^{-6}$ & 50 & 42 \\
         (3.6) & qDLR-PS + RK4 & $L=8$ & $\Delta t=0.027, \varepsilon=10^{-4}$ & 0.44578  & $2.077 \times 10^{-6}$ & 42 & 36 \\
         (3.7) & qDLR-PS + RK4 & $L=8$ & $\Delta t=0.027, \varepsilon=10^{-3}$ & 0.44579  & $3.984 \times 10^{-5}$ & 28 & 22  \\
         (3.8) & qDLR-PS + RK4 & $L=5$ &  Fourier & 0.44496 & $2.017 \times 10^{-6}$ & 42 & 40 \\
         (4.1) & qDLR-AP + RK4 & L=5 & -- & 0.44496 & $1.3091 \times 10^{-5}$ & 118 & 88 \\
         (4.2) & qDLR-AP + RK4 & L=8 & -- & 0.44498 & $3.6562 \times 10^{-5}$ & 436 & 152 \\
         (4.3) & qDLR-AP + RK4 & L=5 & Fourier & 0.44496 & $1.29142 \times 10^{-5}$ & 104 & 104 \\
    \end{tabular}
    \caption{Results for the Whistler wave calculations, where the theoretical oscillation frequency is 0.445279 and the amplitude of the oscillation should remain constant over time. The electron probability amplitude ($g_e$) is the longest tensor train and also has the highest rank. The maximum QTT rank of $g_e$ over all simulation time and the maximum QTT rank at the last time step are reported in the last two columns. (*) The RK4 calculation was not run to completion due to its high ranks. Unless specified otherwise, the truncation threshold used is $\varepsilon = 2.47 \times 10^{-5}$ and a CFL-limited time step is used. Also note that the simulation length only allows one to compute the frequency roughly within two significant figures.}
    \label{tab:whistler}
\end{table}




\subsection{Radiating dipole in 2-dimensional space via Maxwell's equations}
Now consider the simulation of a radiating dipole by solving Maxwell's equations, 
\begin{align}
 \frac{\partial \textbf{B}}{\partial t} &= -\nabla \times \textbf{E}, 
 \label{eq:dBdt} \\
 \frac{\partial \textbf{E}}{\partial t} &= c^2 \nabla \times \textbf{B} - \frac{1}{\epsilon_0} \textbf{J},
 \label{eq:dEdt}
\end{align}
where $\textbf{E}$ denotes the electric fields, $\textbf{B}$ denotes the magnetic fields, $\textbf{J}$ denotes the current density, $c$ is the speed of light, and $\epsilon_0$ is the vacuum permittivity. For simplicity, the units are chosen such that $c=\epsilon_0 = 1$.
%
%
%
%
A dipole in 2-dimensional (2-D) space is generated with the current density $\textbf{J}$ 
\begin{align}
    \textbf{J} = J_0 \, b(x,a) b(y,a) \sin(\omega t) \hat{z}
\end{align}
where the $J_0$ is the strength of oscillation and $b$ is used to represent a unit box of finite width,
\[ b(x,a) = \begin{dcases}
    1 & \text{for } |x| < a/2
    \\
    0 & \text{o.w}
\end{dcases}\]
In the case where $a$ is much smaller than the simulation domain $L_x$, one obtains a farfield radiation pattern of the form $f(r) \sim \text{Re}(\exp(ikr))$, where $k=\omega / c$ is the wavenumber or spatial frequency of the oscillation and $r=\sqrt{x^2 +y^2}$. 
In the following examples, the parameters $a=2\pi/128$, $\omega=4\pi$, and $L_x=10\lambda$ are used, where $\lambda=2\pi c/\omega$. The simulation is run until the fields reach the boundaries of the domain so that implementing a perfectly matched layer or absorbing boundary conditions is not required.

\begin{figure}
    \includegraphics[width=0.5\linewidth, trim={0cm 1cm 0cm 0cm}]{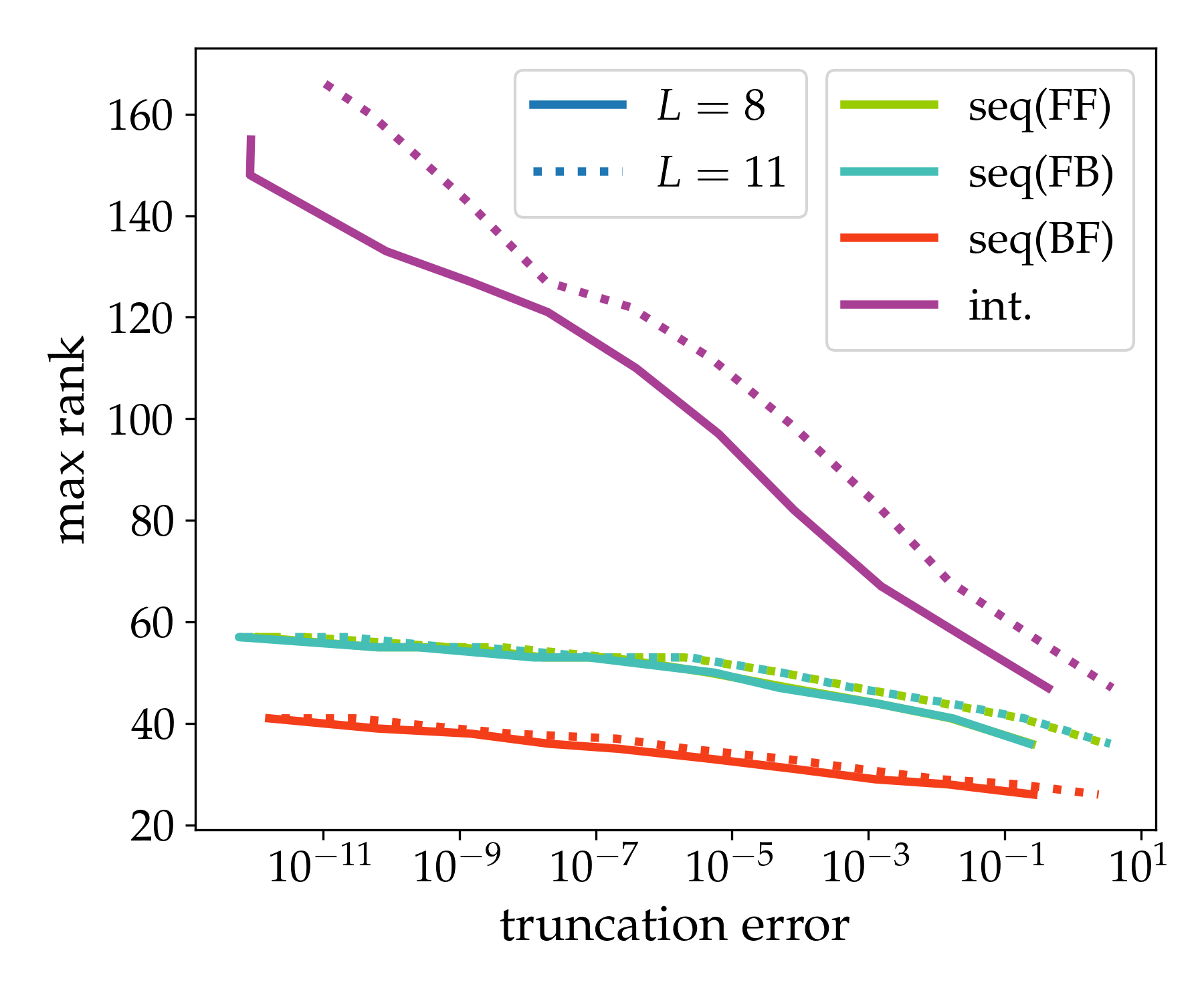}
    \includegraphics[width=0.5\linewidth,  trim={0cm 1cm 0cm 0cm}]{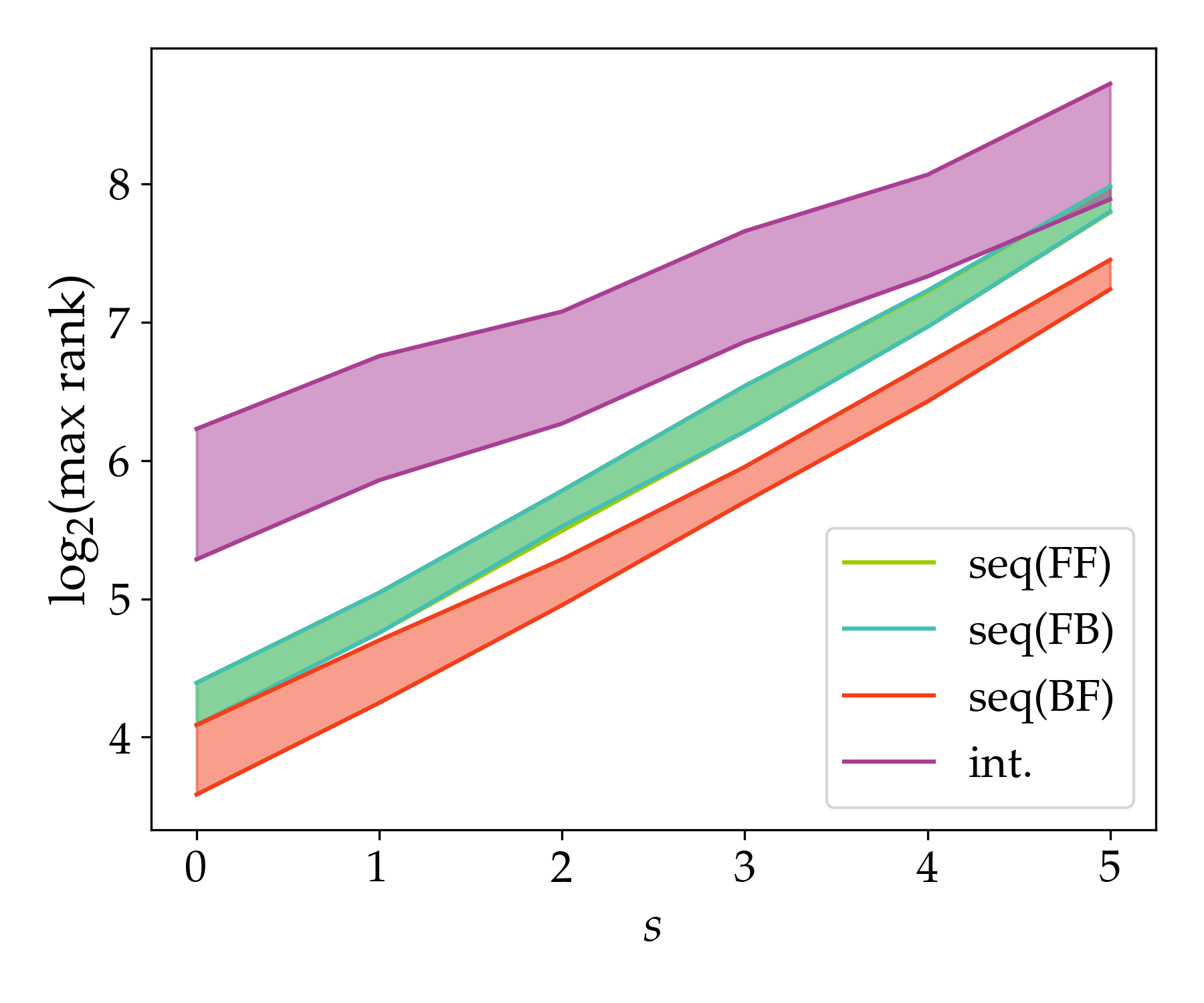}
    \caption{Compressibility of a radial wave $\text{Re}\left[\exp(-ik\sqrt{x^2 +y^2})\right]$ for $k=20$ with $2^L$ grid points along each dimension spanning the domain $x,y\in [-\pi,\pi)$. (left) Ranks as a function of $L_2$ error of the truncated result obtained by performing compression with a specified cutoff tolerance. The seq(BF) mapping yields the lowest ranks, while the interleaved ordering yields the highest ranks. Increasing grid resolution $L$ does not significantly change the ranks, though the $L_2$ errors resulting from truncation for a given $\varepsilon$ are slightly larger, which is within expectation~\cite{Oseledets2011}. (right) QTT ranks as the wavenumber is scaled like $k=20 \times 2^s$ for fixed grid resolution $L=11$. The shaded regions are bounded by the ranks obtained when using a truncation threshold of $\varepsilon=10^{-12}$ and $10^{-6}$. In both plots, the results for seq(FB) and seq(FF) overlap. Together, these results demonstrate that the rank of the radial wave is determined by the spatial frequency of the wave, as opposed to the resolution of the underlying grid, and is relatively high rank. 
    }
    \label{fig:osc_rank}
\end{figure}

In Cartesian coordinates, the radiation pattern is not low-rank. This is within expectation. While plane waves are low-rank in Cartesian coordinates, a radial wave is a superposition of plane waves propagating outward in every direction. Fig.~\ref{fig:osc_rank}(a) shows how the maximum rank of the QTT representation for a radial wave of fixed $k$ varies with the truncation threshold $\varepsilon$. Increasing the resolution (over-resolving the field pattern) does not significantly alter the rank of the QTT. The seq(BF) mapping yields the lowest ranks, while other sequential mappings yield comparable results. The interleaved mapping requires the largest rank. (See Section~\ref{sec:mapping} for definitions of these mappings).  
Fig.~\ref{fig:osc_rank}(b) shows how the rank scales as one increases $k$ on a fixed grid, for an accuracy between $10^{-6}$ and $10^{-12}$. To maintain the same number of grid points per wavelength, one would need to double the number of grid points along each dimension as one doubles $k$. In contrast, the rank increases by less than a factor of two. While this appears to suggest that the QTT representation may be more efficient at extremely high frequencies, it fails to account for the finite accuracy of discretized representation.
Additionally, even though the required rank is less than its maximum possible value of $2^L$, the $r^3$ computational overhead associated with QTT algorithms means that QTTs will likely only have an advantage over a standard dense representation of radial waves on a highly over-resolved grid.


As shown in the previous example, the choice in time integration scheme and the accumulation of numerical errors can affect the rank of the numerical simulation. 
In this section, five different time integration schemes were considered: 
\begin{itemize}
    \item finite difference time domain (FDTD): implemented using step-and-truncate, with rank truncation performed after each Euler time step. The time step size is determined according to the CFL condition.
    \item first-order finite volume (FV): implemented using step-and-truncate, with rank truncation performed after each Euler time step. The time step size is determined according to the CFL condition.
    \item Crank-Nicolson (CN): implemented by using an 1-site alternating least-squares solver \cite{Oseledets2012solution}. A time step size of $\Delta t=0.015$ is used.
    \item Projection-based qDLR, using CN to solve the local problems on each tensor core (qDLR-AP + CN). A time step size of $\Delta t=0.015$ is used.
    \item Single-site qDLR-PS with basis augmentation prior to each sweep, using CN to solve the dynamics of each tensor core (qDLR-PS + CN aug.) with a time step size of $\Delta t=0.015$. First-order projector splitting is used. The basis augmentation is performed by augmenting vector $f$ with $\partial_t f$, such that the TT manifold accurately captures both \cite{yang2020time}.
\end{itemize}
Numerical simulations with fixed rank of $r_{max}=32$ are performed, and the $L_2$ error of the field patterns at the final time step $(t\approx 5.0)$ is measured with respect to the analogous dense calculation,
\begin{align}
    \epsilon^2 = \sum_c \frac{1}{\lVert F_{c, \, dense} \rVert^2} \left \lVert F_{c, \, QTT} - F_{c, \,dense} \right \rVert^2
    \label{eq:EM2_err}
\end{align}
where $F_c$ represents each component of the electric and magnetic fields. (The qDLR calculations are compared to a dense CN calculation.)

\begin{figure}
    \includegraphics[width=0.45\linewidth, trim={0cm 0.5cm 0cm 0cm},]{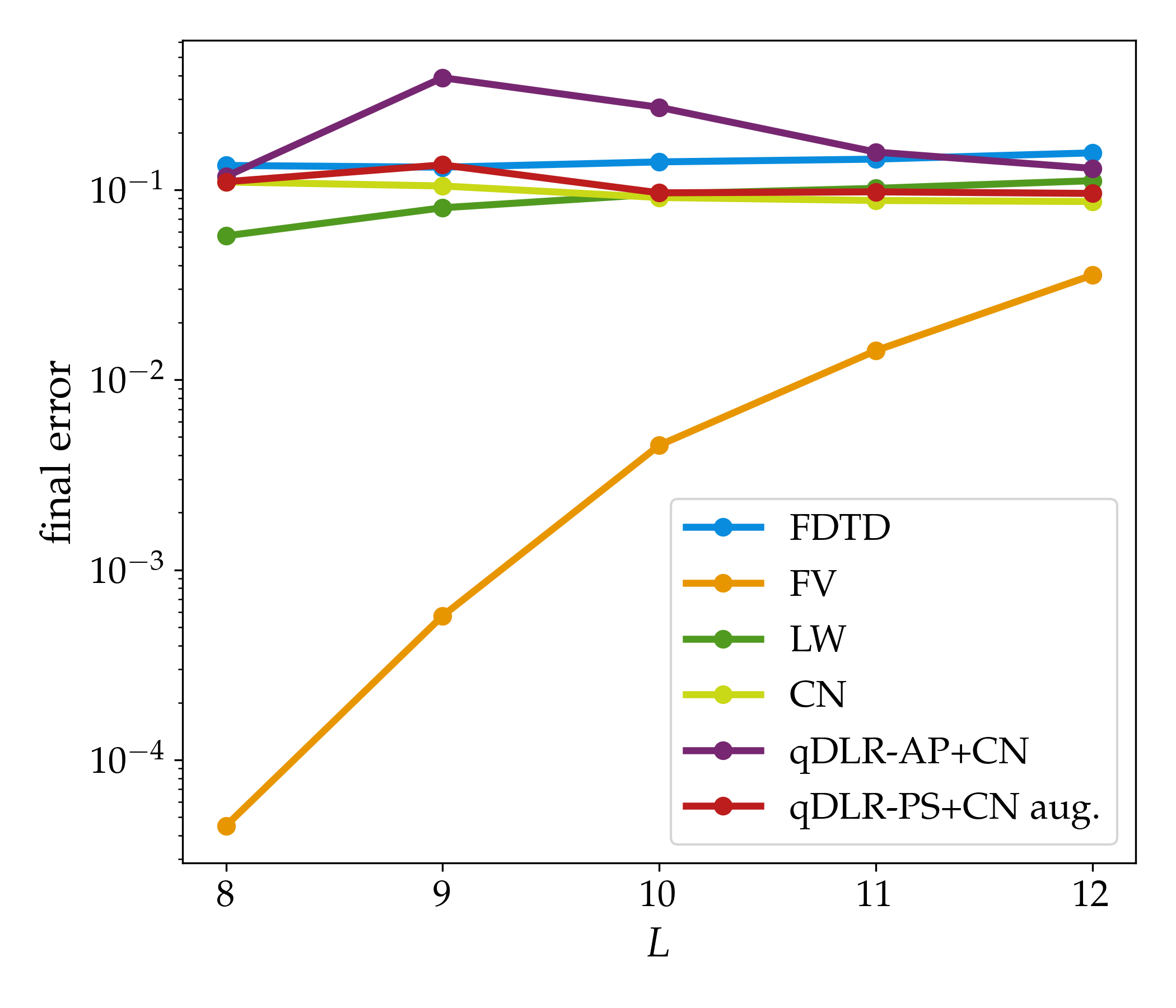}
    \includegraphics[width=0.55\linewidth, trim={0cm 2cm 0cm 0cm},]{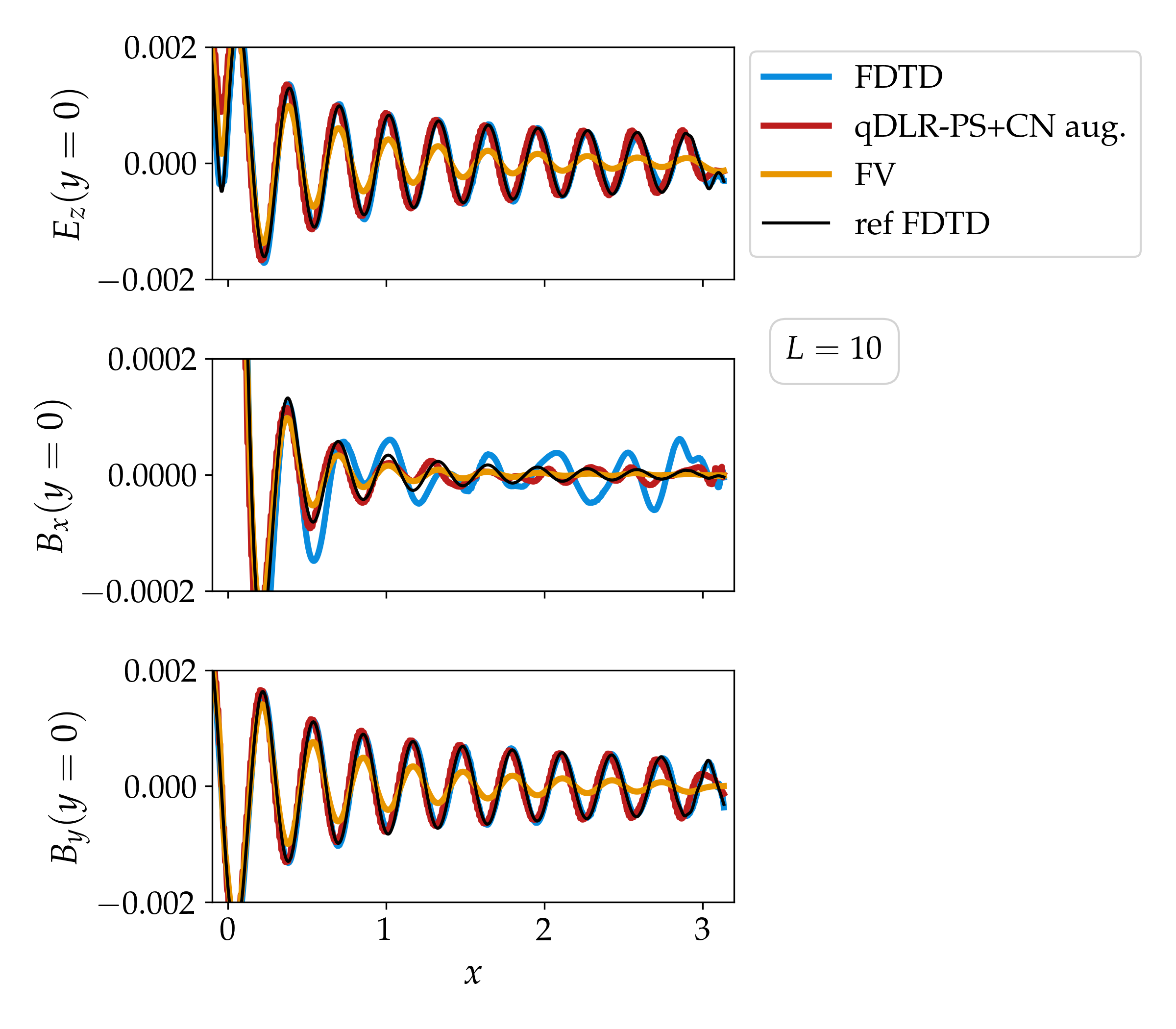}
    \caption{Plots depicting the accuracy of fixed rank QTT calculations for the 2-D radiating dipole. These results were obtained using the seq(BF) mapping. (left) The L$_2$ error of the QTT result with respect to an analogous dense calculation (see Eq.~\ref{eq:EM2_err}) at the final time step, measured for different resolutions $L$. The errors of the qDLR calculations are compared with respect to a dense CN calculation. All time integration schemes use a maximum rank of $r_{max}=32$. (right) Cross sections at $y=0$ of the $E_z$, $B_x$ and $B_y$ fields at the final time step ($t\approx 5.0$). The calculation was performed with $L=10$ calculation and the specified time integration schemes.
    }
    \label{fig:EM2_L_rank_geom}
\end{figure}

\begin{table}[b]
    \centering
    \small
    \begin{tabular}{|c|c|c|c|}
        \hline
         Integrator &  Average time per step (s) & Internal Ranks ($E_z$, $B_x$, $B_y$) \\ 
         \hline
         FDTD & 0.756 & 98, 65, 65 \\
         LW & 3.33 & 131, 98, 98 \\
         FV & 1.49 & 164, 99, 99 \\
         CN & 69.48 & 167, 167, 167 \\
         qDLR-AP + CN & 9.86 & 64, 64, 64 \\
         qDLR-PS + CN (aug.) & 30.15 & 162, 96, 96 \\
         \hline
    \end{tabular}
    \caption{Runtime per time step (averaged over the last ten time steps when the rank is largest) and QTT ranks before the final compression step for 2-D dipole calculations with resolution $L=12$. 
    Note that the code is not optimized and the reported runtimes are only intended to provide a relative sketch of the computational cost.
    }
    \label{tab:EM2D_timings}
\end{table}

The results for the seq(BF) mapping and fixed rank $r_{max}=32$ are shown in Fig.~\ref{fig:EM2_L_rank_geom}, and relative runtimes are reported in Table~\ref{tab:EM2D_timings}. Different QTT mappings showed similar results, aside from the magnitude of the error (see the Section 2.3 of the SI for more results). 
The FV calculation yielded the smallest relative error, but the error grew steeply as resolution is increased. This is due to the overly dissipative nature of the first-order finite volume scheme, yielding lower-rank field patterns at low resolutions. Only at very high resolutions are the field patterns visually comparable to those obtained by FDTD. For similar reasons, the LW calculation showed the same behavior of the error increasing as resolution is increased. However, the error is much larger due to the amount of numerical dissipation being smaller.
In contrast, the other time integration schemes yielded errors that roughly remained constant or even decreased slightly with increasing $L$. 
Note that for qDLR-PS, basis augmentation of the QTT at the beginning of each qDLR sweep was critical for obtaining the reported accuracy. 
However, basis augmentation did make the calculation more costly. While qDLR-PS it did not offer any advantage over FDTD or LW when the time step sizes were comparable, it did provide an avenue for performing implicit time integration and led to reduced overall runtime for the high resolution calculations.

The difference between the FDTD and FV results suggests that incorporating artificial dissipation might be a means of reducing the required rank in a controllable manner. 
%
With artificial dissipation, the equations of motions for the $j^\text{th}$ magnetic and electric field components are 
\begin{align}
 \frac{\partial B_j}{\partial t} &= -(\nabla \times \textbf{E})_j + (-1)^{m/2+1} \sum_{i\neq j} \eta_i \Delta x^m_i \frac{\partial^m}{\partial x_i^m} B_j, \\
 \frac{\partial E_j}{\partial t} &= c^2 (\nabla \times \textbf{B})_j - \frac{1}{\varepsilon_0} J_j + (-1)^{m/2+1} \sum_{i\neq j} \eta_i \Delta x^m_i \frac{\partial^m}{\partial x_i^m} E_j,
\end{align}
where $m$ is an even integer value that denotes the order of the derivative and $\eta_i \geq 0$ controls the strength of the dissipation. $\Delta x_i$ is the grid spacing along the $i^\text{th}$ coordinate. 
In the standard implementation, the second derivatives are approximated with a second-order centered finite-difference stencil. (See SI Section 2.1 for the exact discretized expression). Typically, one also requires $\eta_i \Delta t \leq 1/m^2$ for numerical stability and performance in forward time integrators~\cite{durran:ch3}. 
One approaches the first-order finite volume method when $m=2$ and $\eta_i = \eta_{FV} \equiv \frac{c}{2 \Delta x_i}$. However, because the artificial dissipation is incorporated into the FDTD calculations by simply adding the dissipative term as the fields are updated (before rank truncation), the result is not equivalent to the FV method since it is computed on a staggered space-time grid.
%

\begin{figure}
    \includegraphics[width=0.5\linewidth, trim={0 1cm 0 0}]{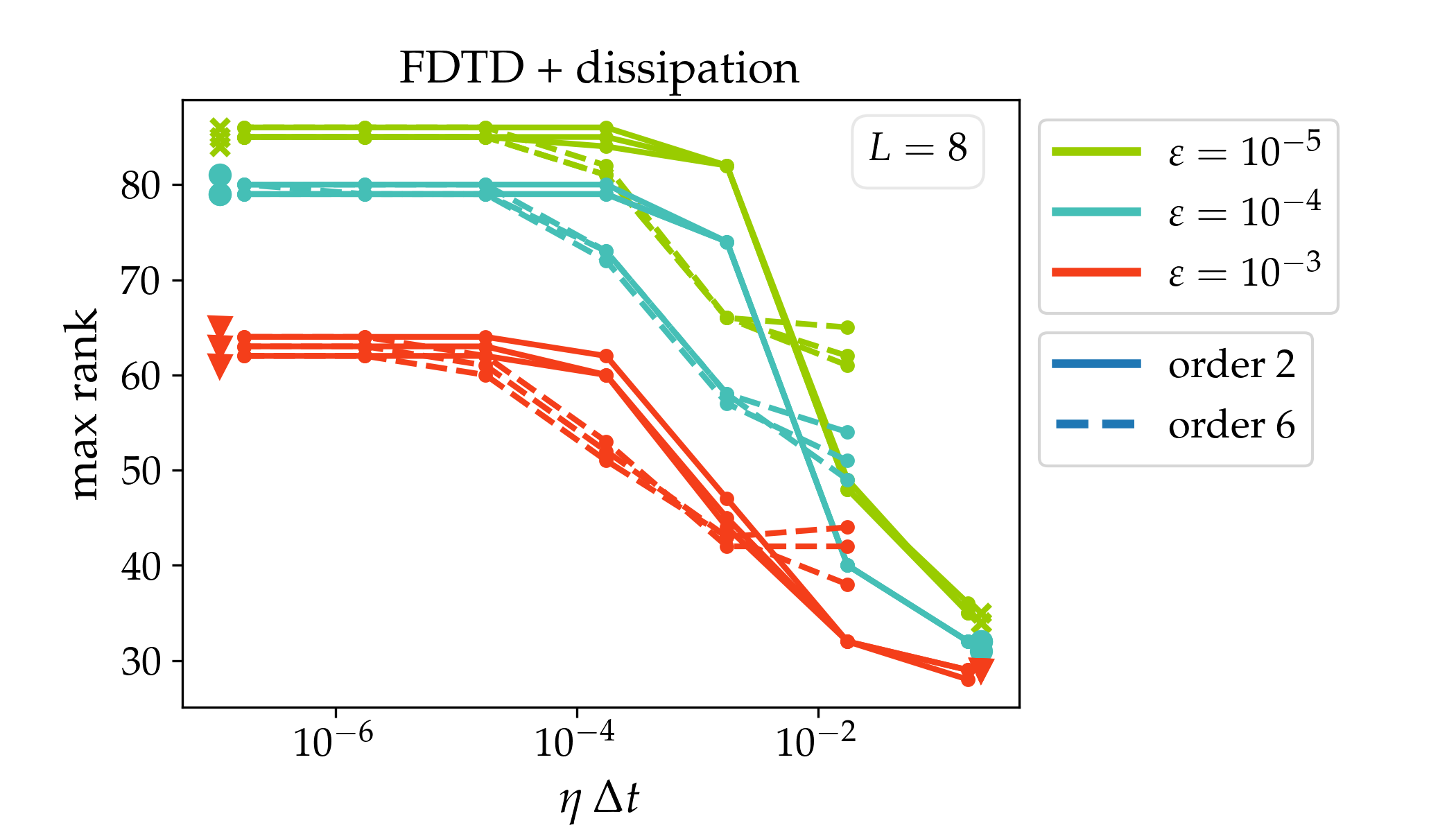}
    \includegraphics[width=0.5\linewidth, trim={0 1cm 0 0}]{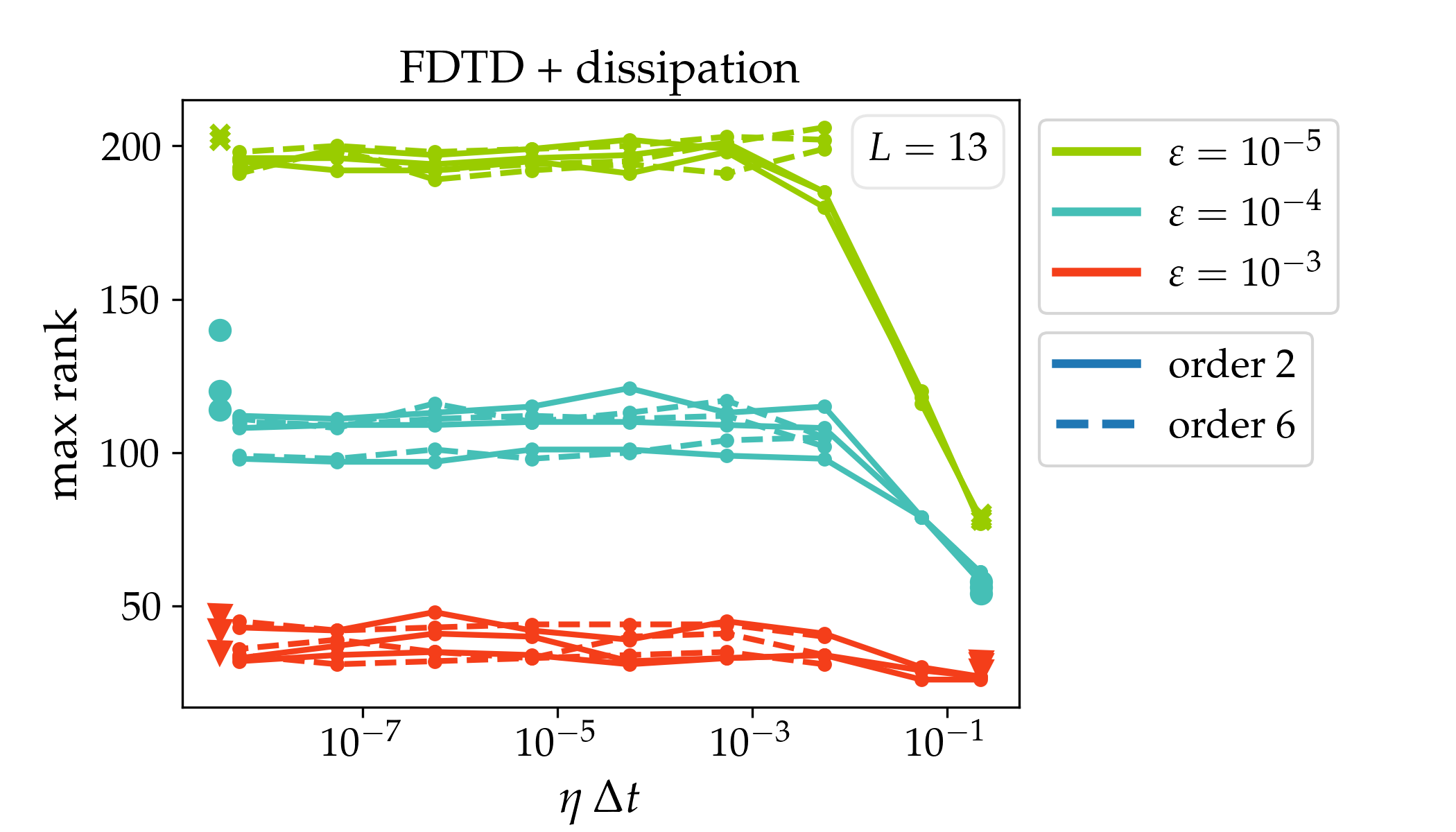}
    \caption{Maximum rank of the field components at the final time step as a function of dissipation strength $\eta$ for an (a) $L=8$ and (b) $L=13$ FDTD calculation. Calculations were performed with the seq(BF) mapping. The large markers on the left of the plots denote the rank required for the FDTD calculation ($\eta=0$), while the large markers on the right denote the rank required for the FV calculation ($\eta=\eta_{FV}$).
    }
    \label{fig:EM2D_eta}
\end{figure}

Fig.~\ref{fig:EM2D_eta} demonstrates how the QTT rank of FDTD calculations varies with respect to $\eta$ for two different grid resolutions ($2^8$ and $2^{13}$ points per dimension) using different truncation thresholds $\varepsilon$.
The value of $\eta$ at which the rank begins to decrease is mostly independent of the truncation threshold $\varepsilon$. It instead depends more on the resolution of the grid. For the $L=8$ calculation, noticeable reduction in rank begins at $\eta \, \Delta t \approx 10^{-4}$ for $m=6$, and $\eta \, \Delta t\approx 10^{-3}$ for $m=2$. However, for the $L=13$ calculation, the rank remains mostly flat except at the largest values of $\eta$ where the rank drops sharply. There is little advantage in trying to tune the numerical dissipation, and one might as well use a finite volume scheme. At such a high resolution, the numerical dissipation is sufficiently small so meaningful results can be obtained anyways. 
These results suggests that the reduction in rank may mostly be a result of mitigating effects arising from using a finite discretization.

\begin{figure}
    \centering
    \subfloat[]{\includegraphics[width=0.45\linewidth, trim={0 0.5cm 0.5cm 0}]{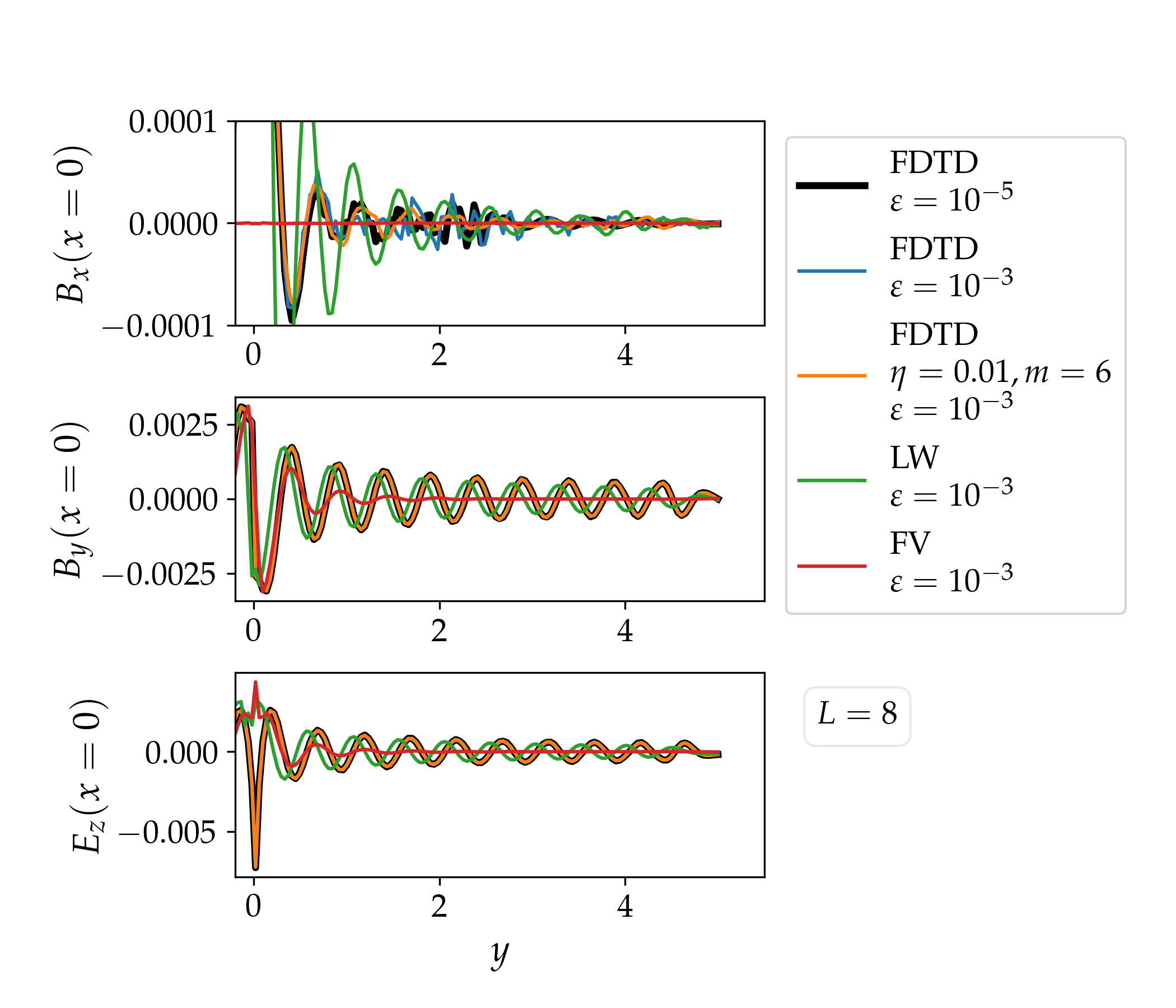}}
    \hspace{0.1cm}
    \subfloat[]{\includegraphics[width=0.53\linewidth, trim={1cm 1cm 0.5cm 0}, clip]{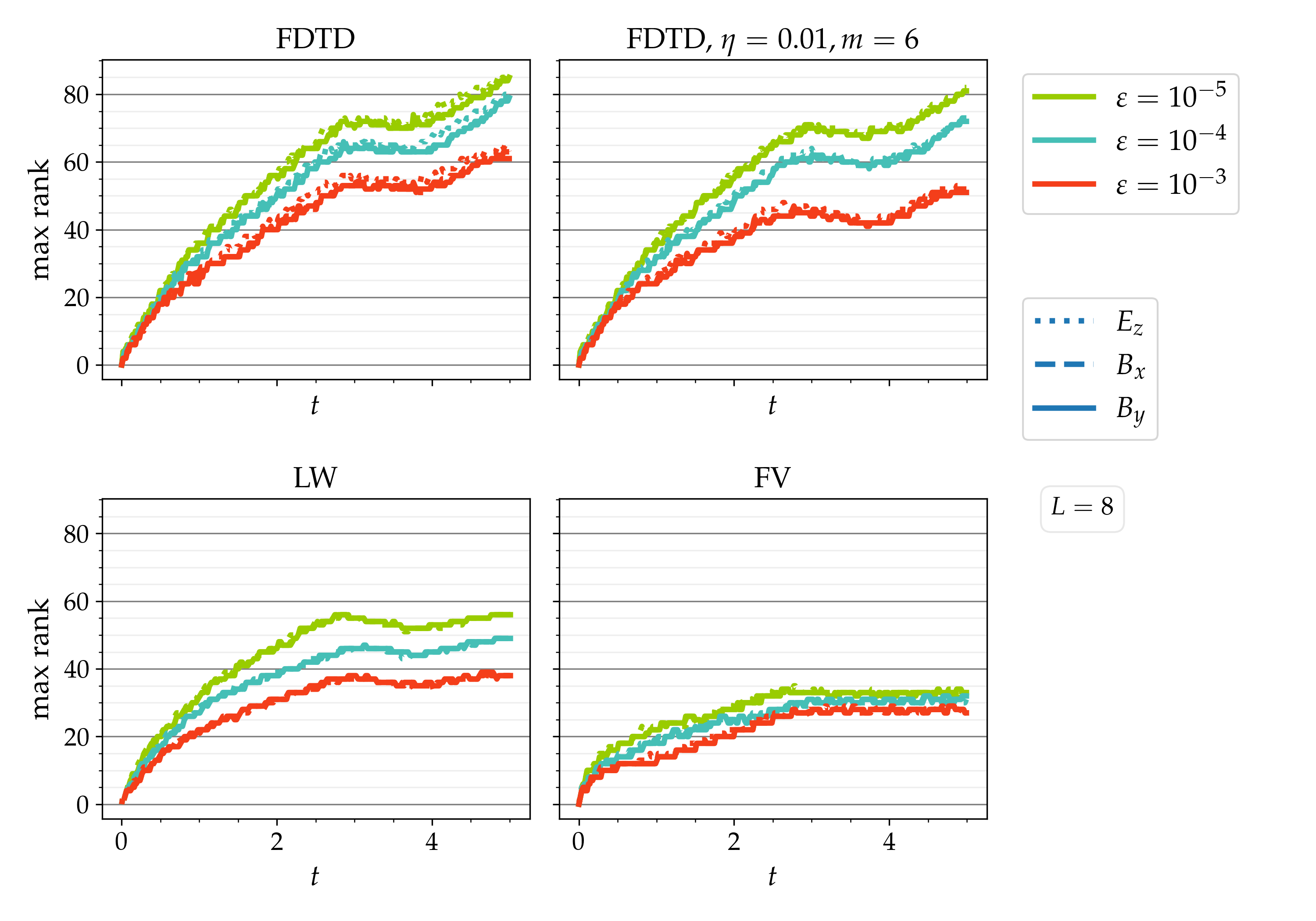}}
    \caption{Results for the 2-D radiating dipole with the seq(BF) mapping with artificial dissipation with $L=8$ with a time step size of $\Delta t \approx  0.0176$. (a) Cross section of field patterns ($B_x$, $B_y$, and $E_z$ at $x=0$) at time $t=5.0$. (b) Maximum QTT rank over simulation time for FDTD, Lax-Wendroff (LW), and first-order finite volume (FV) calculations, as labeled. Calculations were performed with the seq(BF) mapping.
    }
    \label{fig:EM2D_SFBF_L8}
\end{figure}

\begin{figure}
    \centering
    \includegraphics[width=0.45\linewidth, trim={0 0.5cm 0.5cm 0}]{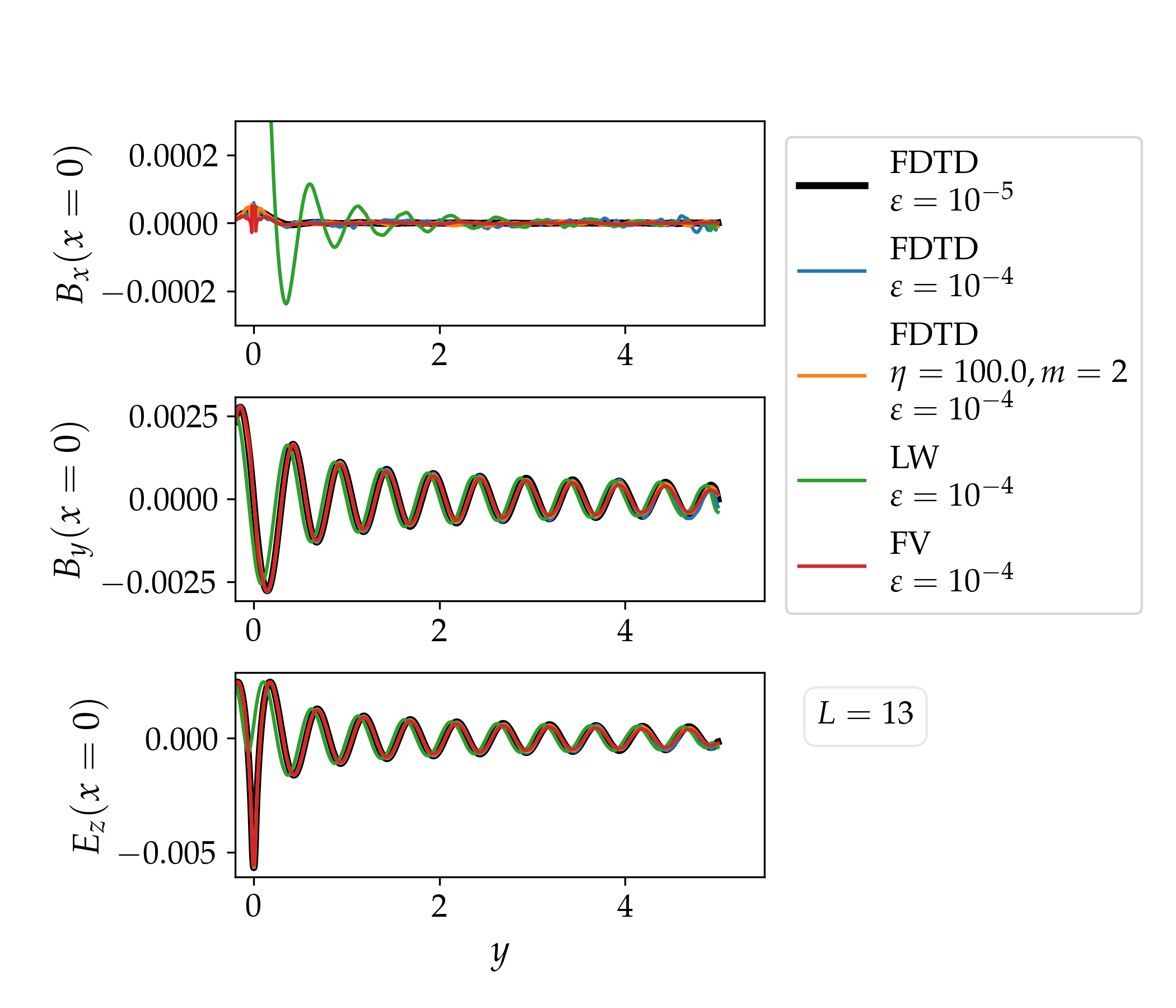}
    \hspace{0.1cm}
    \includegraphics[width=0.53\linewidth, trim={1cm 1cm 0.5cm 0}, clip]{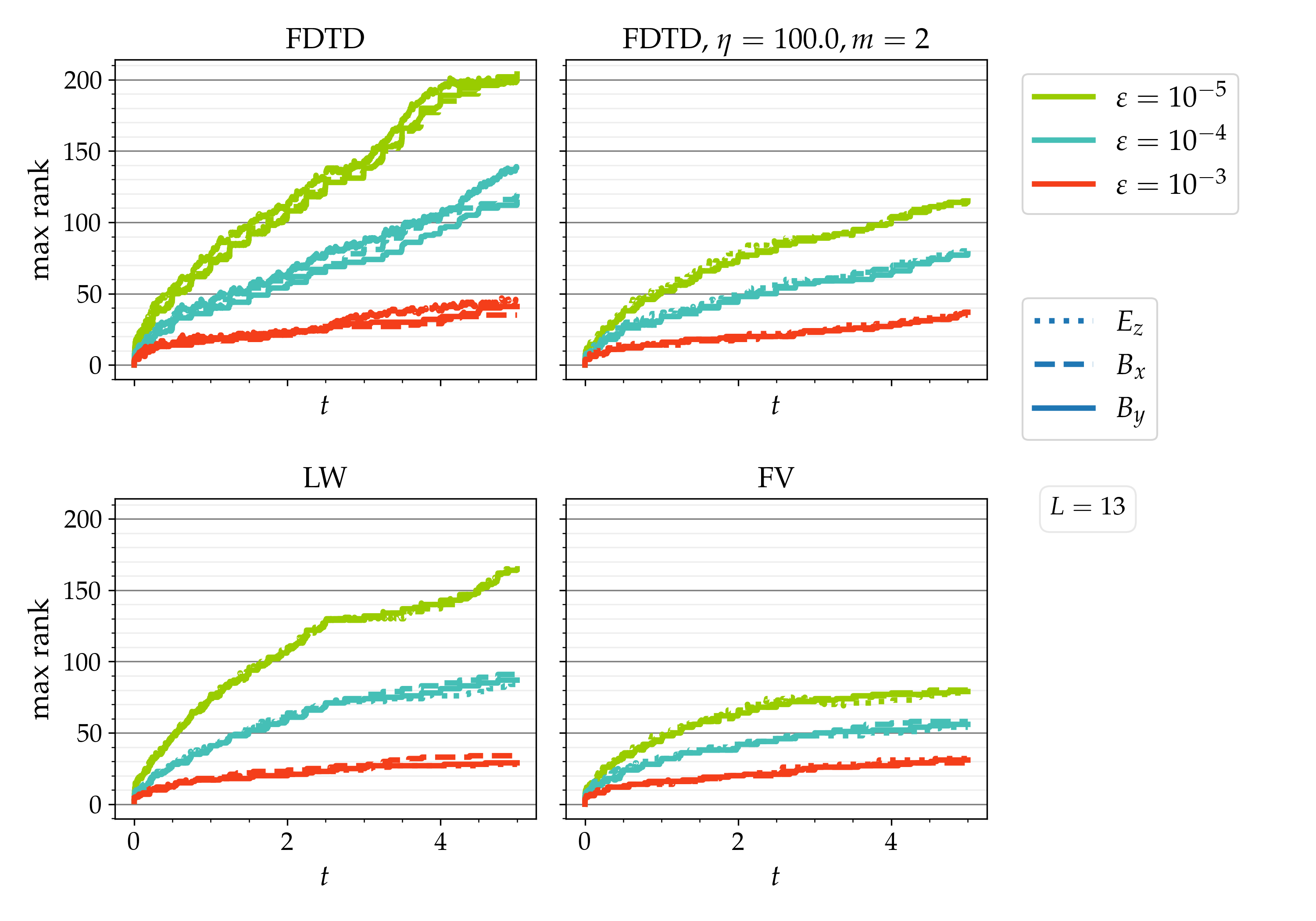}
    \caption{Same as Fig.~\ref{fig:EM2D_SFBF_L8} but for a grid resolution of $L=13$ and time step size $\Delta t \approx 0.00055$.
    }
    \label{fig:EM2D_SFBF_L13}
\end{figure}

Cross-sections of the computed field patterns and the ranks of the field components over time are shown in Fig.~\ref{fig:EM2D_SFBF_L8} for $L=8$ and Fig.~\ref{fig:EM2D_SFBF_L13} for $L=13$. With sufficiently large dissipation, the growth in rank slows over time as the wave propagates outwards, almost plateauing to a moderate rank at late times. Otherwise, the rank grows quickly throughout the simulation, especially if a tighter $\varepsilon$ is used. However, the added dissipation is still able to smooth out some of the more jagged features in the fields that arise from the low-rank approximation.
Results for alternative QTT mappings are provided in the SI (Section 2.3) and exhibit a similar behavior.

\subsection{Radiating dipole in three-dimensional space via the vector potential formulation}

\begin{figure}
    \centering
    \subfloat[]{\includegraphics[width=0.48\linewidth]{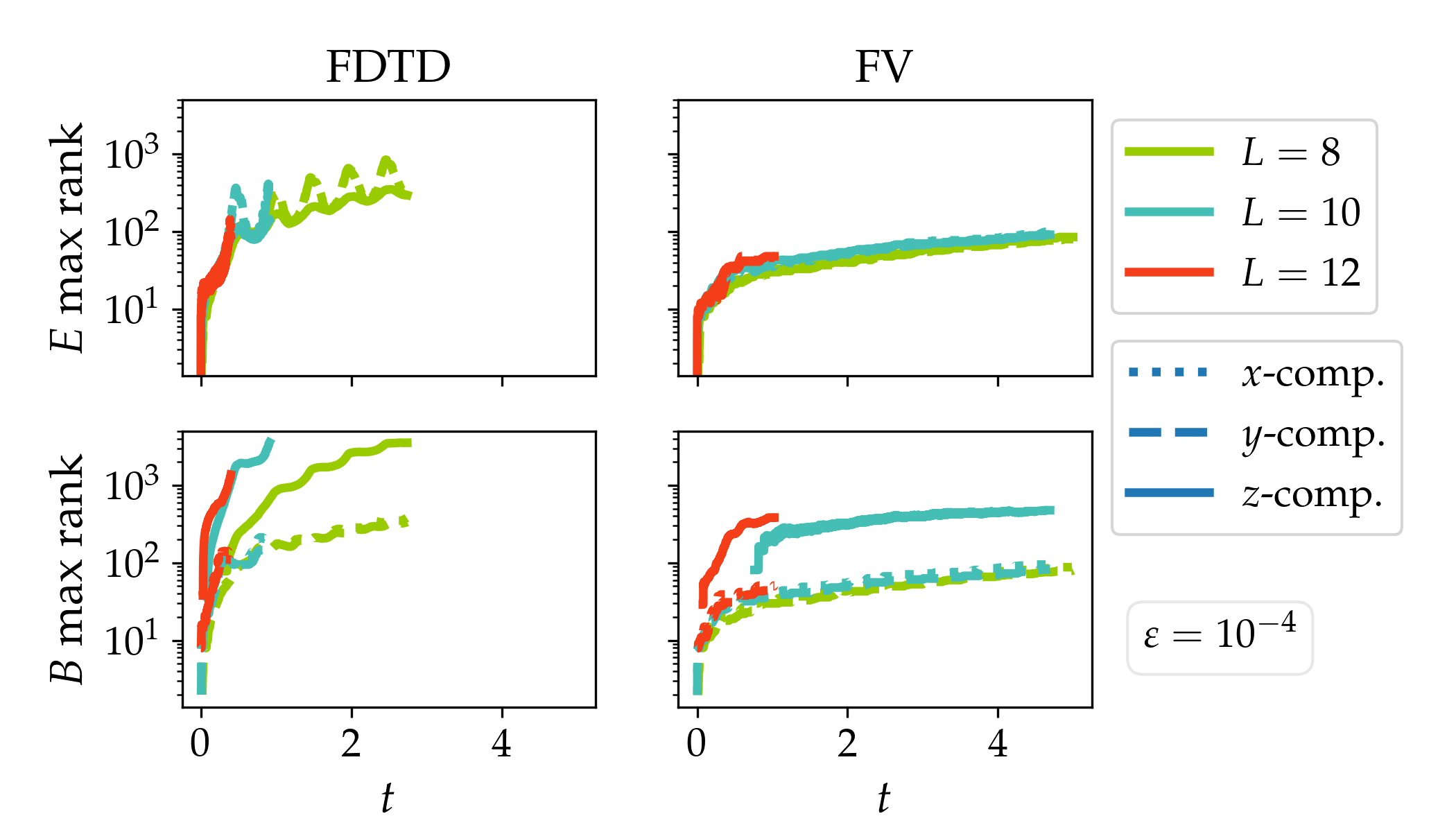}}
    \subfloat[]{\includegraphics[width=0.48\linewidth, trim={0 -1cm 0 0}]{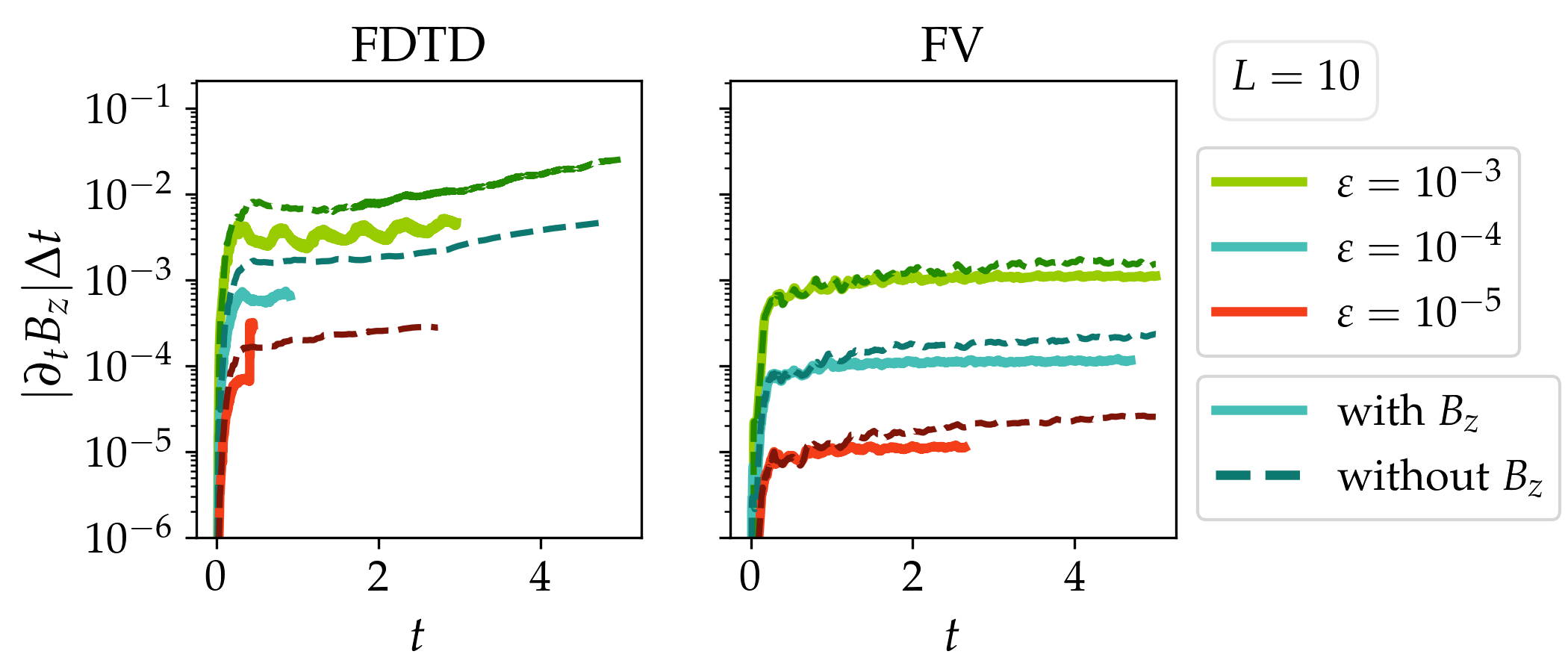}}
    \caption{Results for 3-D dipole radiation solved via Maxwell's equations. (a) Ranks of electric and magnetic fields as one varies resolution. $B_z$ is retained if its $L_2$ norm is greater than $\varepsilon$. The ranks for $B_z$ are significantly larger than the other components, on the order 1000 or larger. Many of the calculations were not finished due to prohibitive costs. 
    (b) Error measured as the $L_2$ norm of $\frac{dB_z}{dt} \, \dt$ over time for different $\varepsilon$. Darker, thinner, and dashed lines are errors when $B_z$ is neglected. The errors are larger without the $B_z$ field but not by a significant amount.
    }
    \label{fig:EM3D}
\end{figure}

For dipole radiation in three dimensions (3-D), the current source is of a finite length in $\hat{z}$,
\begin{align}
    J_z = I_0 \, b(x,a) b(y,a) b(z,\ell) \sin(\omega t)
\end{align}
where $\ell=a$ is used for simplicity.
Unlike the 2-D case where all field components have similar ranks, it is observed that the $B_z$ component suffers from a particularly large rank. This is because the magnitude of the field should be zero due to the symmetry between the $x$ and $y$ coordinates. 
However, after performing the standard rank truncation procedure, there is no guarantee that this constraint will be satisfied. Instead, the $B_z$ component will mostly consist of noise that is no longer low-rank. Increasing resolution yields worse results because the numerical noise occurs at higher frequencies, generating larger errors in the derivative. This occurs for both the FDTD and FV calculation, suggesting that numerical dissipation is unable to resolve this issue. These observations are depicted in Fig.~\ref{fig:EM3D}(a).
Note that this issue is separate from the issue that the divergence-free constraint is not satisfied in the FV method and after making the low-rank approximation.

To mitigate this issue, one could remove fields whose norms are below $\varepsilon$. However, since the error accumulates over time, at some point, the error eventually breaches this threshold. This occurs earlier on in the higher resolution calculations. 
Alternatively, one could completely ignore the field that is expected to be zero and let the error accumulate. 
Fig.~\ref{fig:EM3D}(b) plots the $L_2$ norm of $\frac{dB_z}{dt}$ when the calculation is performed with and without $B_z$, as well as the ranks for these calculations. Because this value should be zero in the ideal case, it can be used as a measure of error. While completely ignoring $B_z$ does increase this error, it is not a significant increase over the error arising from rank truncation.
Still, neglecting $B_z$ is not an ideal solution and this issue demonstrates one pitfall of low-rank methods. 

Instead, one can perform the simulation using the vector potential formulation. In the Lorenz gauge, the equations of motion are
\begin{align}
    & \frac{1}{c^2} \ddt \phi = - \nabla \cdot \textbf{A}, \quad \text{(Lorenz gauge)}
    \\
    & \frac{1}{c^2} \ddt^2 \textbf{A} = \nabla^2 \textbf{A} + \mu_0 \textbf{J},
\end{align}
where 
\begin{align}
    \textbf{B} = \nabla \times \textbf{A}, \quad \textbf{E} = -\nabla \phi - \ddt \textbf{A}.
\end{align}
In addition to preserving the $x$-$y$ symmetry discussed above, the vector potential formulation also automatically satisfies the divergence-free constraint $\nabla \cdot \textbf{B}=0$. Furthermore, in the context of low-rank approximation, the vector potential fields are related to the $\textbf{E}, \textbf{B}$ fields by integration, so they may be smoother and of lower rank. 

One can solve the second-order differential equation by using a leapfrog time evolution scheme,
\begin{align}
    & \textbf{A}^{(n+1)} = 2 \textbf{A}^{(n)} - \textbf{A}^{(n-1)} + c^2\dt^2  \left( \nabla^2 \textbf{A}^{(n)} - \mu_0 \textbf{J}^{(n)} \right)
    \\
    & \phi^{(n+1/2)} = \phi^{(n-1/2)} - c^2 \dt (\nabla \cdot \textbf{A}^{(n)})
\end{align}
or by also writing an explicit expression for $\ddt \textbf{A}$,
\begin{align}
    \ddt 
    \begin{bmatrix}
        \textbf{A} \\ \ddt \textbf{A} \\ \phi
    \end{bmatrix}
    = 
    \begin{bmatrix}
        0 & \pmb{I} & 0 \\
        c^2 \nabla^2 & 0 & 0 \\
        -c^2 \nabla \cdot & 0 & 0 
    \end{bmatrix}
    \begin{bmatrix}
        \textbf{A} \\ \ddt \textbf{A} \\ \phi
    \end{bmatrix}
    + 
    \begin{bmatrix}
        0 \\ \frac{1}{\epsilon_0} \textbf{J} \\ 0
    \end{bmatrix}
    \label{eq:Aphi}
\end{align}
and solving directly. Here, the Lorenz gauge is used directly to solve for the time derivative of $\phi$. 
In the 3-D case radiating dipole case, only $A_z$, $\ddt A_z$, and $\phi$ are non-zero. 


\begin{figure}
    \includegraphics[width=\linewidth, trim={0 1.5cm 0 0}]{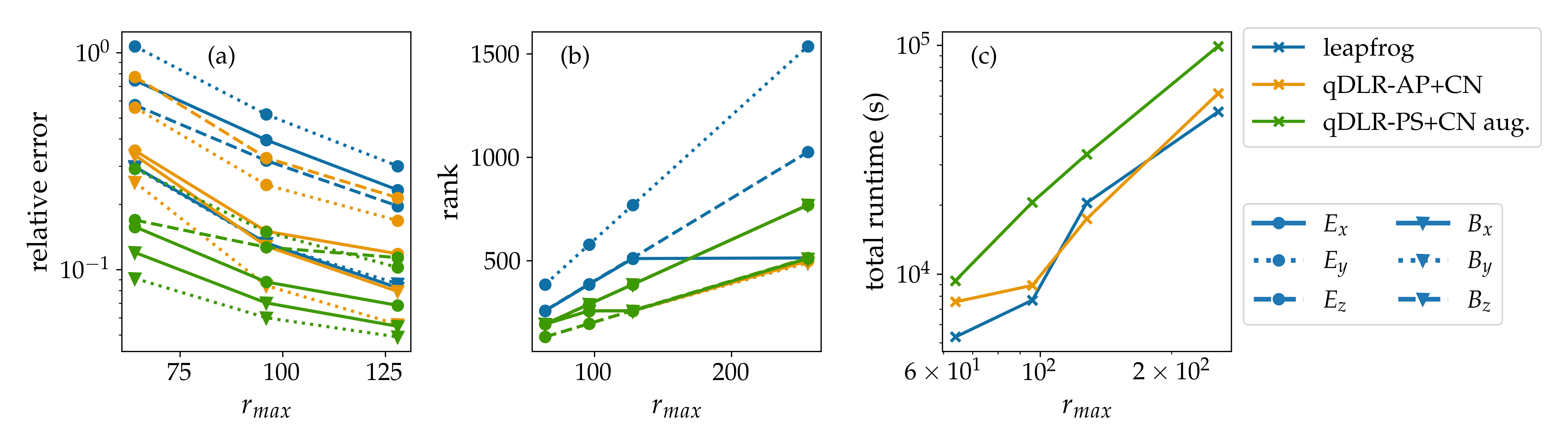}
    \caption{Results for 3-D dipole radiation computed via the vector-potential formulation. Calculations contain $2^{10}$ grid points along each dimension. A seq(FFF) mapping is used. (a) Error of calculations capped to a maximum rank of 64, 96, and 128, measured with respect to a calculation with $r_{max}=256$. Different colors correspond to different time integration schemes, while line textures and line markers are used to denote the different components of the $\textbf{E}$ and $\textbf{B}$ field. Note that no lines show up for $B_z$ since it is exactly zero in all case. 
    (b) Rank of the $\textbf{E}$ and $\textbf{B}$ fields at the final time step once computed from $\textbf{A}$ and $\phi$.
    (c) Approximate total runtimes of the different time integration schemes with respect to rank. 
    }
    \label{fig:Aphi_3D}
\end{figure}

A 3-D calculation with a resolution of $2^{10}$ grid points along each dimension was performed using the leapfrog, qDLR-PS with CN, and qDLR-PS with CN time integrators. Directly performing CN with an ALS-based solver was too expensive with our current implementation. The seq(FFF) mapping with physical dimensions ordered as ($x,y,z$) was used. Because of the cost of the computation, the calculation was performed at different maximum ranks and the error was measured with respect to the results obtained with $r_{max}=256$. Note that even $r_{max}=256$ is lower than the rank one would obtain if only a rank truncation threshold $\varepsilon$ were specified and the rank was allowed to increase as needed. However, the field patterns with $r_{max}=256$ for the different time integration schemes are visually similar, suggesting that the calculations are reasonable. One-dimensional cross-sections of the fields are provided in the SI Section 3.1. 

As shown in Fig.~\ref{fig:Aphi_3D}(a), the error drops slowly as one increases the rank, and However, unlike when solving Maxwell's equations directly, $B_z$ remains zero due to symmetry. Furthermore, one only has to compute the dynamics of three field components as opposed to five or six components. The downside of the vector potential formulation is that the electric and magnetic fields are no longer immediately accessible. However, they can be computed from the potential fields. The ranks of the $\textbf{E}, \textbf{B}$ fields are shown in Fig.~\ref{fig:Aphi_3D}(b). Interestingly, the $\textbf{E}$ field ranks are noticeably larger for the leapfrog calculation. Lastly, Fig.~\ref{fig:Aphi_3D}(c) plots the total runtime for the different time integration schemes. qDLR-PS is takes the longest, but provides the most accurate results. Due to the large numerical errors arising from rank truncation, a first-order splitting scheme is sufficient for qDLR-PS. 


The sequential QTT mapping used here is not expected to be the optimal layout, since any coupling between the $x$-dimension tensors and the $z$-dimension tensors must be carried through all of the $y$-dimension tensors. However, for fixed rank calculations, the seq(FFF) mapping still yields better results compared to the interleaved mapping 
In the future, different quantized tensor network ansatz\"{e} beyond the tensor train should be investigated.

\section{Discussion}

This work demonstrated that the choice in time integrator, the numerical discretization used, and the choice in problem representation can affect the efficiency and performance of QTT calculations. 

Regarding the choice in time integrator, step-and-truncate time integration of low-order methods such as FDTD and the Lax-Wendroff are fairly cheap to implement. Some amount of numerical dissipation, whether inherent to the time integrator or incorporated artificially, appeared to help reduce the growth in rank over simulation time. Higher order methods can also be implemented. However, the more stages the time integration scheme has, the more difficult the calculation becomes, especially if one insists on performing the rank truncation only at the final stage for the best accuracy. This issue could potentially be mitigated by using randomized or sketching methods to perform the rank-truncation procedure, and would require further investigation. 
Time integration with qDLR, especially qDLR-PS, also yielded relatively good results. Both high-order and implicit time integrators can be easily used with qDLR, allowing one to take larger time steps. However, the main disadvantage is that the accuracy of the integrator depends on the span of the QTT manifold that full dynamics is projected onto to obtain the reduced dynamics at each tensor core. Augmenting this basis is critical for improving accuracy, especially when the rank of the QTT manifold is small. Despite the rank increase, the calculation typically still remains feasible. 

Regarding the choice in numerical discretization, it was observed that for some calculations, increasing grid resolution or using a spectral basis could actually help reduce or avoid unexpected growth in QTT rank over simulation time. Note that this was not true in all cases. For example, utilizing RK4 with SAT often led to very large ranks that also increased with resolution.
%
Unfortunately, increasing grid resolution typically reduces the allowable time step size for explicit time integrators. 
Instead of increasing resolution, one could incorporate artificial dissipation to help control the growth in QTT rank. However, adding too much dissipation reduces the fidelity of the simulation. 


Finally, QTTs (and low-rank methods in general) unsurprisingly face difficulties when dealing with zero or near-zero magnitude fields. These often arise from the subtraction of two large magnitude fields. Thus, the rank truncation threshold used on the large magnitude fields must be smaller than the near-zero magnitude field, such that it can be accurately captured. Otherwise, the near-zero magnitude field will quickly be dominated by errors arising from the low-rank approximation and require a large rank. Adding numerical dissipation is generally not sufficient for removing the errors arising from rank truncation. If a field must be exactly zero, it is possible to simply ignore that field and accept the fact that there will be some error associated with this approximation. However, a more satisfactory case would be to formulate the problem such that the constraint is satisfied by construction. In this paper, this was achieved by performing the radiating dipole simulation using the vector potential formalism for Maxwell's equation.


\section{Conclusion}
This paper investigates different QTT time integration schemes for advection-dominated PDEs, as well as possible methods for controlling the QTT rank during the simulation. This is important for keeping the costs of the QTT numerical simulation low. The performance and ranks of the QTTs highly depend on the time integrator used. For step-and-truncate time integrators, the QTT ranks are most controlled if a dissipative numerical scheme is used. Without numerical dissipation, a higher resolution or a spectral basis should be used to avoid grid discretization errors from causing rank increase after long times. For quantized dynamical low-rank time integrators, there is more flexibility in which time integrator to use. It appears that their primary advantage over step-and-truncate methods is the ability to easily perform implicit time integration. 

\section{Code Availability}
Code can be made available upon request.

\section{Acknowledgments}
EY thanks Y.L. for proofing the initial draft. 

\section{Funding}
This material is based in part upon work supported by the Laboratory Directed Research and Development Program of Lawrence Berkeley National Laboratory under U.S. Department of Energy Contract No. DE-AC02-05CH11231, and in part upon work supported by the U.S. Department of Energy, Office of Science, Office of Advanced Scientific Computing Research's Applied Mathematics Competitive Portfolios program under Contract No. AC02-05CH11231.
This research used resources of the National Energy Research Scientific Computing Center, a DOE Office of Science User Facility supported by the Office of Science of the U.S. Department of Energy under Contract No. DE-AC02-05CH11231 using NERSC award ASCR-ERCAPm1027.

\appendix

\section{Whistler wave initialization}
\label{app:whistler}
In the test case, the magnetic field is oriented along the $x$-direction $\textbf{B}_0 = B_0 \hat{x}$, and has a perturbation in the electron drift velocity, 
\begin{align}
    u_{e,y} &= A \cos(kx - \omega t) = \frac{A}{2} \left( e^{ikx} + e^{-ikx} \right)
    \\
    u_{e,z} &= -A \sin(kz - \omega t) = \frac{A}{2} \left( ie^{ikx} - i e^{-ikx} \right)
\end{align}
We use the following ansatz\"{e} for the current, electric field, and magnetic field: 
\begin{align}
    \textbf{J} &= \hat{y} J_y \cos(kx-\omega t) + \hat{z} J_z \sin(kx-\omega t)
    \\
    \textbf{E} &= \hat{y} E_y \sin(kx-\omega t) + \hat{z} E_z \cos(kx-\omega t)
    \\
    \textbf{B} &= \hat{y} B_y \cos(kx-\omega t) + \hat{z} B_z \sin(kx-\omega t)
\end{align}
Plugging into Maxwell's equations, we obtain
\begin{align}
    E_y &= \left(1 - \frac{c^2 k^2}{\omega^2} \right)^{-1} \frac{1}{\epsilon_0 \omega} J_y
    \\
    E_z &= -\left(1 - \frac{c^2 k^2}{\omega^2} \right)^{-1} \frac{1}{\epsilon_0 \omega} J_z
    \\
    B_y &= -\frac{k}{\omega} E_z
    \\
    B_z &= \frac{k}{\omega} E_y
\end{align}

For the calculations in the main text, the units are normalized such that $\varepsilon_0=1$, $c=1$, and $B_0=1$. Additionally, the electron mass $m_e$, skin depth $\delta_e$, and cyclotron frequency $\omega_{c,e}$ are all equal to one, and the electron thermal velocity is $v_{th,e}=10^{-3}$. The dynamics of the ions are neglected. Waves with wavevector $k=1.0 d_e^{-1}$ are considered.

\printbibliography

\includepdf[pages=-]{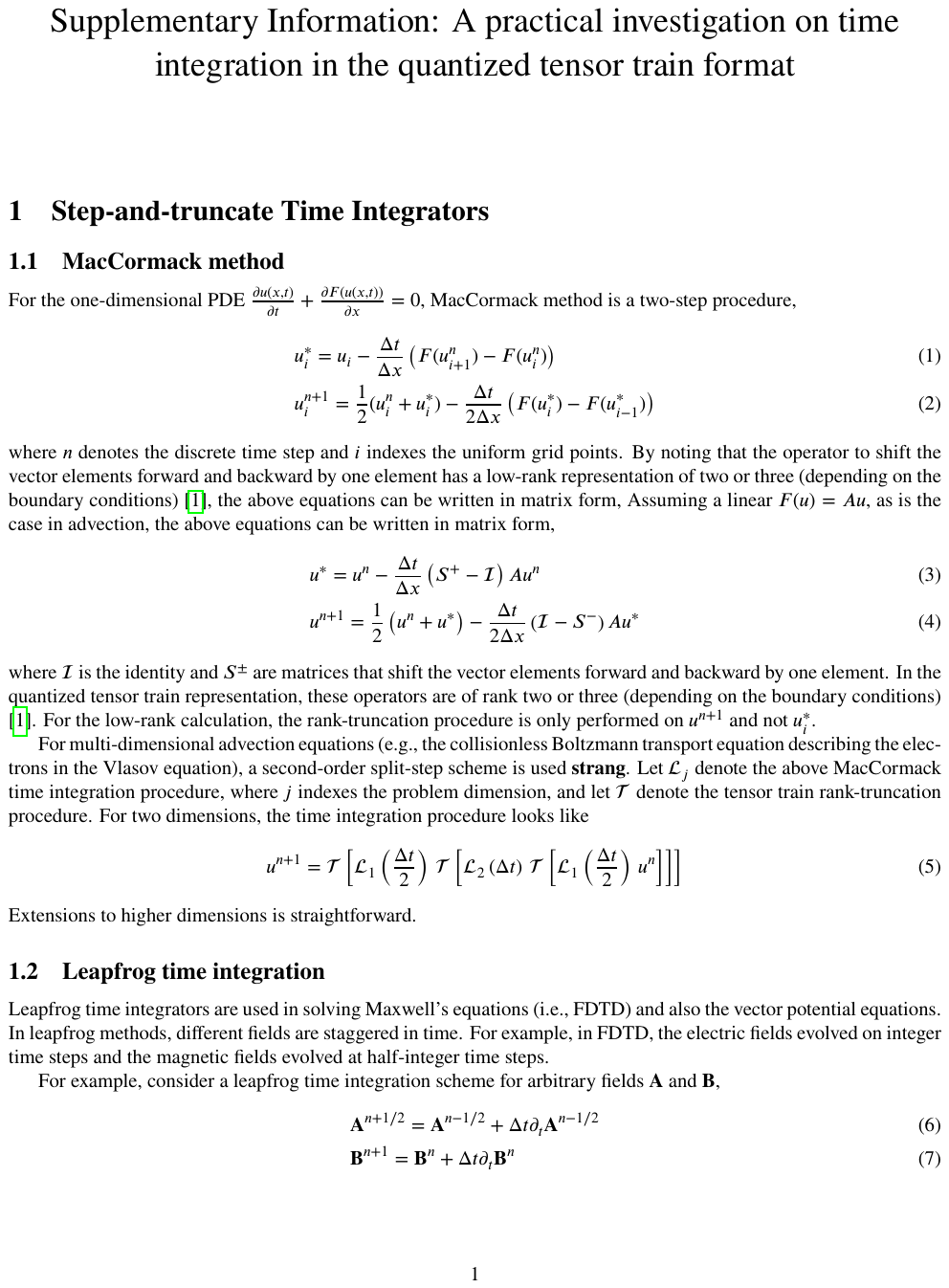}

\end{document}